\documentclass[twocolumn]{aastex701}

\usepackage{natbib}
\usepackage{multirow}
\usepackage{graphicx}
\usepackage{amsmath}
\usepackage[normalem]{ulem}
\usepackage{mathrsfs}
\usepackage{threeparttable}
\usepackage{booktabs}
\usepackage{array}
\usepackage{bm}
\usepackage{xspace}
\usepackage{subcaption}
\usepackage{hyperref}
\usepackage{float}

\usepackage{silence}
\newcommand{\degree}{$^{\circ}$\xspace}
\newcommand{\cm}{cm$^{-3}$\xspace}
\newcommand{\cmtwo}{cm$^{-2}$\xspace}

\newcommand{\msun}{M$_{\odot}$\xspace}
\newcommand{\kms}{km~s$^{-1}$\xspace}

\newcommand{\peryr}{yr$^{-1}$\xspace}
\newcommand{\persec}{s$^{-1}$\xspace}

\newcommand{\jykms}{Jy~km~s$^{-1}$\xspace}
\newcommand{\mjyb}{mJy~beam$^{-1}$\xspace}
\newcommand{\Mjysr}{MJy~sr$^{-1}$\xspace}

\newcommand{\h}{$^{\mathrm{h}}$\xspace}
\newcommand{\m}{$^{\mathrm{m}}$\xspace}
\newcommand{\s}{$^{\mathrm{s}}$\xspace}

\newcommand{\htwoSfive}{H$_2$~S(5)\xspace}

\newcommand{\htwo}{H$_2$\xspace}
\newcommand{\hfivea}{H5$\alpha$\xspace}
\newcommand{\hii}{\ion{H}{2}\xspace}
\newcommand{\hi}{\ion{H}{1}\xspace}
\newcommand{\arii}{[\ion{Ar}{2}]\xspace}
\newcommand{\neii}{[\ion{Ne}{2}]\xspace}
\newcommand{\COthreetwo}{CO(3--2)\xspace}
\newcommand{\COtwone}{CO(2--1)\xspace}
\newcommand{\COonezero}{CO(1--0)\xspace}
\newcommand{\COsixfive}{CO(6--5)\xspace}

\accepted{September 12, 2026}
\submitjournal{AJ}
\shortauthors{Sheriff et al.}

\begin{document}

\title{JWST Observations of Starbursts: A Young Bubble in NGC 253's Central Starburst}

\correspondingauthor{Kaitlyn E. Sheriff}

\author[0009-0005-8382-0614]{Kaitlyn E. Sheriff}
\affiliation{Department of Physics and Astronomy, University of Kansas, 1251 Wescoe Hall Drive, Lawrence, KS 66045, USA}
\email[show]{kaitlyn.sheriff@ku.edu}

\author[0000-0001-8782-1992]{Elisabeth A.C. Mills}
\affiliation{Department of Physics and Astronomy, University of Kansas, 1251 Wescoe Hall Drive, Lawrence, KS 66045, USA}
\email{eacmills@ku.edu}

\author[0000-0002-5582-8521]{Xinyu Mai}
\affiliation{Department of Physics and Astronomy, University of Kansas, 1251 Wescoe Hall Drive, Lawrence, KS 66045, USA}
\email{xinyu-mai@ku.edu}

\author[0000-0003-3633-0098]{Utsav Siwakoti}
\affiliation{Department of Physics and Astronomy, University of Kansas, 1251 Wescoe Hall Drive, Lawrence, KS 66045, USA}
\email{siwakotiutsav@ku.edu}

\author[0000-0003-0014-0508]{Sara E. Duval}
\affil{Ritter Astrophysical Research Center, University of Toledo, Toledo, OH 43606, USA}
\email{saraduval018@gmail.com}

\author[0000-0002-5480-5686]{Alberto D. Bolatto}
\affiliation{Department of Astronomy, University of Maryland, College Park, MD 20742, USA}
\affiliation{Joint Space-Science Institute, University of Maryland, College Park, MD 20742, USA}
\email{bolatto@umd.edu}

\author[0000-0003-2508-2586]{Rebecca C. Levy}
\affiliation{Space Telescope Science Institute, 3700 San Martin Drive, Baltimore, MD 21218, USA}
\email{rlevy.astro@gmail.com}

\author[0000-0001-7863-5047]{Jia Wei Teh}
\affiliation{Universit\"{a}t Heidelberg, Interdisziplin\"{a}res Zentrum f\"{u}r Wissenschaftliches Rechnen, Im Neuenheimer Feld 225, 69120 Heidelberg, Germany}
\email{jiaweiteh.astro@gmail.com}


\author[0000-0002-5666-7782]{Torsten B\"oker}
\affiliation{European Space Agency, c/o STScI, 3700 San Martin Drive, Baltimore, MD 21218, USA}
\email{boeker@stsci.edu}

\author[0000-0002-9511-1330]{Serena A. Cronin}
\affiliation{Department of Astronomy, University of Maryland, College Park, MD 20742, USA}
\email{cronin@umd.edu}

\author[0000-0002-5782-9093]{Daniel~A.~Dale}
\affiliation{Department of Physics and Astronomy, University of Wyoming, Laramie, WY 82071, USA}
\email{ddale@uwyo.edu}

\author[0009-0004-5807-9142]{Keaton Donaghue}
\affiliation{Department of Astronomy, University of Maryland, College Park, MD 20742, USA}
\email{}

\author[0000-0001-6527-6954]{Kimberly Emig}
\affil{National Radio Astronomy Observatory, 520 Edgemont Road, Charlottesville, VA 22903, USA}
\email{kemig@nrao.edu}

\author[0000-0001-6708-1317]{Simon C.~O.\ Glover}
\affiliation{Universit\"{a}t Heidelberg, Zentrum f\"{u}r Astronomie, Institut f\"{u}r Theoretische Astrophysik, Albert-Ueberle-Str.\ 2, 69120 Heidelberg, Germany}
\email{}

\author[0000-0002-2775-0595]{Rodrigo Herrera-Camus}
\affiliation{Departamento de Astronomía, Universidad de Concepción, Concepción, Chile}
\affiliation{Millennium Nucleus for Galaxies, MINGAL}
\email{rhc@udec.cl}

\author[0000-0002-0560-3172]{Ralf S.\ Klessen}
\affiliation{Universit\"{a}t Heidelberg, Zentrum f\"{u}r Astronomie, Institut f\"{u}r Theoretische Astrophysik, Albert-Ueberle-Str.\ 2, 69120 Heidelberg, Germany}
\affiliation{Universit\"{a}t Heidelberg, Interdisziplin\"{a}res Zentrum f\"{u}r Wissenschaftliches Rechnen, Im Neuenheimer Feld 225, 69120 Heidelberg, Germany}
\email{}

\author[0000-0001-8490-6632]{Thomas S.-Y. Lai}
\affil{IPAC, California Institute of Technology, 1200 East California Boulevard, Pasadena, CA 91125, USA}
\email[]{ThomasLai.astro@gmail.com}

\author[0000-0002-1254-4174]{Ashley E. Lieber}
\affiliation{Department of Physics and Astronomy, University of Kansas, 1251 Wescoe Hall Drive, Lawrence, KS 66045, USA}
\email{ashleylieber@ku.edu}

\author[0000-0003-4023-8657]{Laura Lenki\'{c}}
\affiliation{IPAC, California Institute of Technology, 1200 East California Boulevard, Pasadena, CA 91125, USA}
\email{laura.lenkic@gmail.com}

\author[0000-0002-2644-0077]{Sebastian Lopez}
\affiliation{Department of Astronomy, The Ohio State University, 140 W. 18th Ave., Columbus, OH 43210, USA}
\affiliation{Center for Cosmology and AstroParticle Physics, The Ohio State University, 191 W. Woodruff Ave., Columbus, OH 43210, USA}
\email{lopez.764@osu.edu}

\author[0000-0001-9436-9471]{David S. Meier}
\affiliation{New Mexico Institute of Mining and Technology, 801 Leroy Place, Socorro, NM, 87801, USA}
\email{}

\author[0000-0001-8224-1956]{J\"{u}ergen Ott}
\affiliation{National Radio Astronomy Observatory, PO Box O, 1003 Lopezville Road, Socorro, New Mexico 87801, USA}
\email{jott@nrao.edu}

\author[0000-0003-1545-5078]{J.D.T. Smith}
\affil{Ritter Astrophysical Research Center, University of Toledo, Toledo, OH 43606, USA}
\email{JD.Smith@utoledo.edu}

\author[0000-0003-4209-1599]{Yu-Hsuan Teng}
\affiliation{Department of Astronomy, University of Maryland, College Park, MD 20742, USA}
\email{yhteng@umd.edu}

\author[0000-0001-5434-5942]{Paul P.~van der Werf}
\affiliation{Leiden Observatory, Leiden University, PO Box 9513, 2300 RA Leiden, The Netherlands}
\email{pvdwerf@strw.leidenuniv.nl}

\author[0000-0002-3158-6820]{Sylvain Veilleux}
\affiliation{Department of Astronomy, University of Maryland, College Park, MD 20742, USA}
\affiliation{Joint Space-Science Institute, University of Maryland, College Park, MD 20742, USA}
\email{}

\author[0000-0002-5877-379X]{Vicente Villanueva}
\affiliation{Instituto de Estudios Astrofísicos, Facultad de Ingeniería y Ciencias, Universidad Diego Portales, Av. Ejército Libertador 441, 8370191 Santiago, Chile}
\affiliation{Millennium Nucleus for Galaxies, MINGAL}
\email{vicente.avl365@gmail.com}


\author[]{Rachel Cionitti}
\affiliation{Department of Physics and Astronomy, University of Kansas, 1251 Wescoe Hall Drive, Lawrence, KS 66045, USA}
\email{r964c568@ku.edu}

\author[0009-0006-9141-4381]{Joseph Havens}
\affiliation{Department of Physics and Astronomy, University of Kansas, 1251 Wescoe Hall Drive, Lawrence, KS 66045, USA}
\email{joe.havens79@ku.edu}

\author[0009-0008-9238-5871]{Maleah Rhem}
\affiliation{Department of Physics and Astronomy, University of Kansas, 1251 Wescoe Hall Drive, Lawrence, KS 66045, USA}
\email{mkrhem@ku.edu}

\author[0009-0003-7692-9764]{Nathan Shaw}
\affiliation{Department of Physics and Astronomy, University of Kansas, 1251 Wescoe Hall Drive, Lawrence, KS 66045, USA}
\email{nshaw@ku.edu}

\author[0009-0002-1465-1958]{Kai Smith}
\affiliation{Department of Physics and Astronomy, University of Kansas, 1251 Wescoe Hall Drive, Lawrence, KS 66045, USA}
\email{smith.kai682@ku.edu}

\author[0009-0007-0155-8312]{Hazel B. Wright}
\affiliation{Department of Physics and Astronomy, University of Kansas, 1251 Wescoe Hall Drive, Lawrence, KS 66045, USA}
\email{hbwright@ku.edu}

\author[0000-0002-3696-2127]{Md Abdullah Al Zaman}
\affiliation{Department of Physics and Astronomy, University of Kansas, 1251 Wescoe Hall Drive, Lawrence, KS 66045, USA}
\email{abdullah.alzaman@ku.edu}


\begin{abstract}

\noindent We present a multi-wavelength analysis of a young bubble in the nuclear starburst of NGC~253 using new JWST MIRI–MRS observations together with archival ALMA (100, 350, 690~GHz) and Chandra data. The MIRI maps reveal a prominent bubble-like structure in both ionized and molecular emission lines. The bubble is spatially coincident with one of the least embedded massive young clusters detected with ALMA, suggesting that the cluster is driving the expansion. 
We measure a radius of $\sim$ 11.5 $\pm$ 3.4~pc and an expansion velocity of $\sim$ 90 $\pm$ 44~\kms, implying a dynamical age of $\sim$ 0.1 $\pm$ 0.1~Myr. Using RADEX modeling of multiple CO transitions, we infer a molecular mass in the range of $(1.3 \pm 0.3) \times 10^4$ to $(2.8 \pm 0.8) \times 10^5$~\msun.
We derive a kinetic energy on the order $10^{51-52}$~erg, consistent with mechanical input from Wolf-Rayet stellar winds or supernovae in a $\sim 10^{6}$~\msun cluster. The existence of a large population of Wolf-Rayet stars or past supernovae is supported by the presence of coincident X-ray emission. Our results provide direct evidence that individual clusters in a nuclear environment can carve out coherent structures on parsec scales and inject significant energy and momentum into the surrounding interstellar medium, which can contribute to the nuclear outflow in NGC~253.

\end{abstract}

\keywords{
\uat{Interstellar medium}{847};
\uat{Starburst galaxies}{1570};
\uat{Young star clusters}{1833};
\uat{Superbubbles}{1656};
\uat{Galaxy winds}{626}
}

\section{Introduction\label{sec:introduction}}


Large-scale outflows driven by both active galactic nuclei and stellar feedback from forming massive star clusters and clustered supernovae (SNe) play a key role in galaxy evolution. Outflows grow from expanding bubbles of hot gas ($T \sim 10^{7}$~K) that sweep the interstellar medium (ISM) into shells of shocked material \citep{Heckman+1990, Thompson+2024}. Expanding bubbles eventually reach a ``blowout" phase, where the shocked gas escapes the disk, enriching the surrounding circumgalactic and intergalactic medium \citep{Oppenheimer&Dave2008, Fielding+2017}, and altering the metal content of the disk \citep{Finlator&Dave2008, Peeples&Shankar2011}. The resulting outflow can often be modeled as a bi-cone, in which the material escapes both above and below the plane of the disk \citep{Cronin+2025}. Some of this outflowing material may later accrete back onto the galaxy, potentially fueling new star formation \citep[e.g.,][]{Heckman+1990, Veilleux+2005, Marasco+2019, Veilleux+2020, Li+2023}.

\begin{figure*}[htbp]
    \centering
    \includegraphics[width = \linewidth]{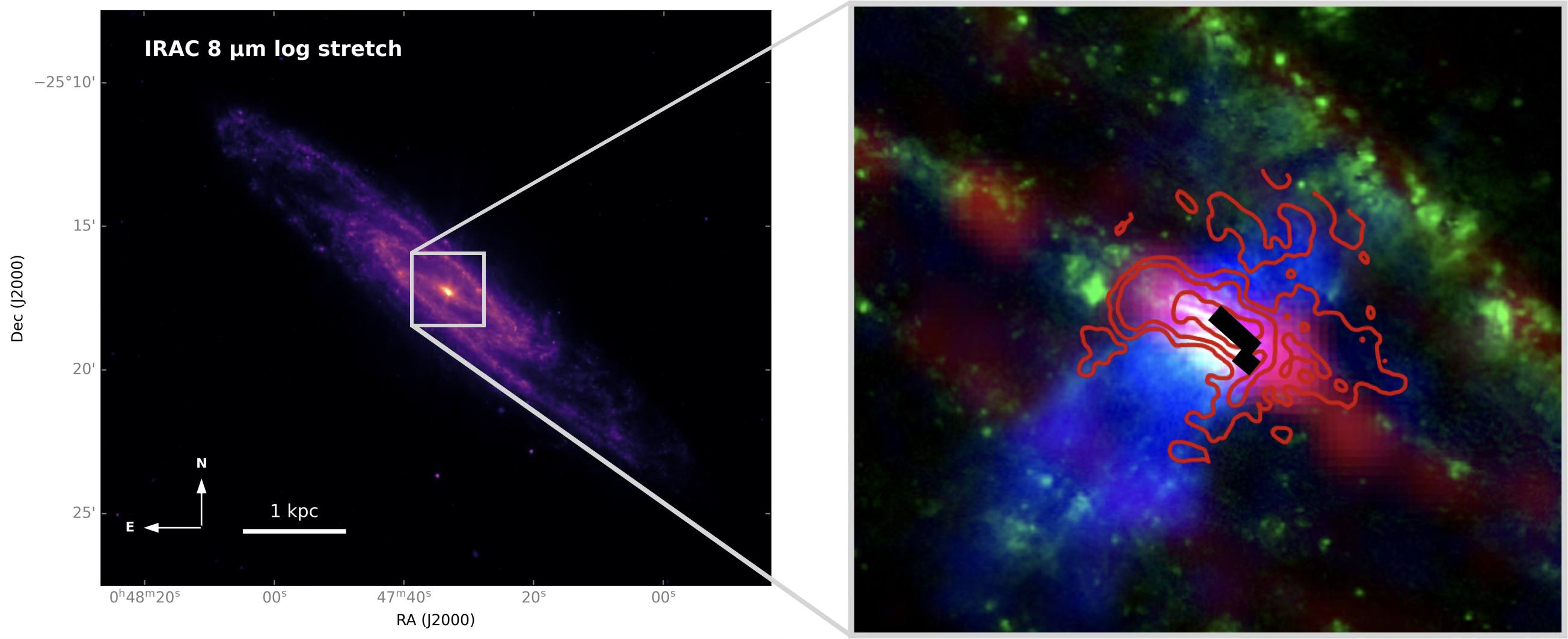}
    \caption{Overview of the multiphase outflow in NGC~253. \textbf{Left:} Spitzer IRAC 8~$\micron$ image from the Local Volume Legacy survey, shown with a logarithmic stretch \citep{Dale+2009, Lee+2009}. The white box marks the central region shown in the right panel, and the scale bar corresponds to 1~kpc. \textbf{Right:} Composite view of the nuclear starburst region highlighting the base of the large-scale superwind. The background three-color image is adapted from \citet{Lopez+2023}, where blue shows broadband (0.5--7~keV) X-ray emission, green shows H$\alpha$ emission \citep{Lehnert+1996}, and red shows \COtwone emission \citep{Leroy+2021}. Red contours indicate the molecular outflow detected in CO \citep{Bolatto+2013}. The JWST footprint for observations presented in this paper is overlaid in solid black.}
    \label{fig:wind_overview}
\end{figure*}

At a distance of 3.5~Mpc \citep{Rekola+2005, Newman+2024, Congiu+2025}, NGC~253 is one of the nearest starburst galaxies which exhibits a large-scale wind \citep[which has been studied across the electromagnetic spectrum; see Figure~\ref{fig:wind_overview}; e.g.,][]{Ulrich+1978, Strickland+2000, Strickland+2002, Westmoquette+2011, Lehmer+2013, Bolatto+2013, Leroy+2015, Krieger+2019, Levy+2021, Levy+2022, Lopez+2023, Cronin+2025}, making it an ideal environment for studying how feedback from intense star formation impacts the surrounding ISM. The center of NGC~253 has been undergoing vigorous star formation for a few $\times 10^7$~years \citep{Rieke+1980, Engelbracht+1998}. Stars in the central half kiloparsec form at a rate of $\sim 2.8$~\msun~\peryr \citep{Ott+2005, Bendo+2015, Leroy+2015}, leading to a high rate of supernova (SN) explosions ($\sim 0.1$~\peryr) \citep{Rieke+1988, vanBuren+1994, Ulvestead+1997}. This rapid injection of energy into the ISM has produced a kiloparsec-scale outflow of hot gas, commonly referred to as a superwind \citep{Ulrich+1978, Tomisaka+1988, Heckman+1990}, which can be easily viewed due to the galaxy's nearly edge-on orientation \citep[$i\approx 78$\degree;][]{Pence1980, Westmoquette+2011}.

A majority of the recent starburst activity in NGC~253's nucleus is occurring in young, massive embedded clusters, or ``super" star clusters \citep[SSCs;][]{Leroy+2018, Mills+2021, Levy+2021, Levy+2022}. Although a few SSCs are observed to have small-scale outflows, the total molecular gas mass of these outflows \citep[$\sim 10^3$--$10^6$~\msun;][]{Levy+2021} is much less than the molecular gas mass of the entire wind \citep[$\sim 2$--$4 \times 10^7$~\msun;][]{Zschaechner+2018}, so it remains unclear how and when the outflows from individual clusters couple to the overall superwind. A main limitation has been the high extinction toward the center of NGC~253 \citep[e.g.,][]{Leroy+2018, Cronin+2025}, which limits observations of highly-ionized tracers in the innermost region of the starburst. New JWST MIRI-MRS observations (as discussed in Section~\ref{sec:jwst_data}) of NGC~253's nucleus give us an opportunity to better compare the ionized gas properties to the molecular gas visible with ALMA, and to identify how far the feedback from the current generation of embedded clusters extends into the surrounding ISM. 

In this paper, we present new JWST and archival ALMA observations of the central starburst region of NGC~253 (Section~\ref{sec:observations}). Section~\ref{sec:results_analysis} details our analysis of the molecular and ionized gas tracers toward a prominent bubble-like source, which was previously detected in \neii\ $12.8~\micron$ emission and identified as a shell by \citet{Boeker+1998} and \citet{Gunthardt+2015}. The latter also identified this source as a candidate nucleus of NGC~253, a claim we revisit in Section~\ref{sec:cluster_xray}. We investigate the morphology, kinematics, and energetics of this structure to assess its connection to feedback from the nuclear starburst. In Section~\ref{sec:Discussion}, we interpret these results in the context of cluster-driven feedback and the galactic superwind. Finally, Section~\ref{sec:Conclusion} summarizes our findings and their implications for star formation and outflow processes in NGC~253.

\section{Observations\label{sec:observations}}

\subsection{JWST Data\label{sec:jwst_data}}

The JWST Cycle 1 GO program \#~1701 (PI: A. Bolatto) carried out observations of NGC~253 and M82 to obtain MIRI and NIRCam imaging of their multiphase outflows and MIRI Medium-Resolution Spectrometer \citep[MRS;][]{Rieke+2015} spectroscopy of their cores. These observations have yielded a series of recent publications focusing on M82 \citep{Levy+2024, Bolatto+2024, Fisher+2025, Villanueva+2025, Lopez+2025, Duval+2026, Cronin+2026}.

In this study, we analyze the MIRI-MRS observations of NGC~253, which were obtained from the Mikulski Archive for Space Telescopes at the Space Telescope Science Institute and can be accessed via \dataset[doi: 10.17909/rvag-ak82]{https://doi.org/10.17909/rvag-ak82}. The MRS consists of four co-spatial integral field units covering 4.9--27.9~$\micron$. Integral field units targeting longer wavelengths have larger footprints, ranging from 3\farcs2 $\times$ 3\farcs7 in Channel 1 up to 6\farcs6 $\times$ 7\farcs7 in Channel 4. Additionally, spectral resolving power decreases as a function of wavelength from $R \approx 3500$ at 5~$\micron$ to $R \approx 1500$ at 28~$\micron$ \citep{Wells+2015, Law+2023}. The analysis in this paper focuses only on Channel 1 (4.9--7.65~$\micron$). Channels 1--4 will be presented in R.~C.~Levy et al.\ in prep (focusing on the ionized gas in the clusters), U.~Siwakoti et al.\ in prep (focusing on extinction), K.~Donaghue et al. in prep. (focusing on the \hi lines), and Y.-H.~Teng et al.\ in prep (focusing on \htwo line modeling toward the clusters).

The MIRI-MRS observations of NGC~253 were taken on 2023-11-13, covering the nuclear starburst, embedded clusters, and the base of the molecular outflow with four pointings (see overlaid footprint in right panel of Figure~\ref{fig:wind_overview}). The total on-source time is 4 hours with an additional background pointing taking 1.17 hours. These data were reduced with all three stages of the JWST Calibration Pipeline using version 1.17.1 and \texttt{jwst\_1322.pmap} for the reference file. Several changes were made to the default pipeline to limit detector artifacts. During the \texttt{straylight} step in the calwebb\_spec2 pipeline, the continuum had been over-subtracted on either side of the bright \arii emission line at 6.99~$\micron$. We substitute the reference file, \texttt{mrsxartcor\_hack.fits} (D. Law 2024, priv comm.), for the \texttt{straylight} step to correct for this issue. Additionally during the calwebb\_spec2 pipeline, the \texttt{residual fringing} step is enabled. Lastly, pixel-based background subtraction is chosen as the method of background subtraction using dedicated background observations. For more details, see R.~C.~Levy et al.\ in prep and \citet{Duval+2026}, in which the same reduction process is followed for both NGC~253 and M82, respectively.

\subsubsection{Data Processing\label{sec:data_processing}}

For our analysis, we focus on bright, short-wavelength Channel 1 lines, which provide the highest spatial resolution (0\farcs35; 5.9~pc) in the MIRI–MRS dataset. We choose representative tracers of both the molecular and ionized gas components allowing us to probe the morphology and kinematics of the bubble and its environment at small scales ($\sim 6$~pc). The MIRI-MRS Channel 1 transitions used in this analysis are listed in Table~\ref{tab:line_properties}.

\begin{table}[htbp]
\centering
\caption{Atomic and molecular transitions considered in this work.}
\label{tab:line_properties}
\begin{tabular}{lccccc}
\hline\hline
Species & Transition & Wavelength & $E_{\rm upper}$ \\
 &  & ($\micron$) & (K) \\
\hline
\arii & 2P$^0$--2P$^0$ 3/2--1/2 & 6.9853 & 2059.3 \\
\hline
\hfivea & 6--5 & 7.4599 & 4384.9 \\
\hline
\htwo & 0--0 S(5) & 6.9095 & 4586.059 \\
\hline
CO & 1--0 & 2603 & 5.5 \\
 & 2--1 & 1301 & 16.6 \\
 & 3--2 & 867 & 33.2 \\
 & 6--5 & 433.8 & 116.2 \\
\hline
\end{tabular}
\end{table}

To further prepare the MIRI-MRS Channel 1 data for analysis post-pipeline reduction, we constructed continuum-subtracted line cubes for the molecular and ionized lines that are analyzed in this paper (see Table~\ref{tab:line_properties}). For each subcube, we extracted an asymmetric velocity range of $(-1000, +1250)$~\kms (based on visual inspection of cubes of the brightest lines, i.e. \arii) around the rest wavelength of each line to account for the systemic velocity of NGC~253 \citep[243~\kms;][]{Koribalski+2004} and to ensure full coverage of asymmetric line profiles. The prominent \arii emission may in part reflect the relatively low ionization potential of Ar (15.8 eV; \citealt{Sansonetti2005}), which allows Ar$^+$ to be readily produced in \hii regions, and not significantly depleted onto dust grains or incorporated into molecules due to its noble-gas nature \citep{PerezMontero+2007}. The continuum was estimated for each line cube by fitting a first-order polynomial to line-free regions immediately adjacent to each line at each spatial pixel. A spatially-varying continuum model was then subtracted from each pixel of the data to isolate the emission lines. 

After continuum subtraction, the cubes were uniformly smoothed and re-gridded to create a consistent spatial and spectral sampling across the dataset. Each image slice of the cube was convolved with a 2D Gaussian kernel chosen to match the point spread function full-width at half-maximum (FWHM) of 0\farcs35 derived by \citet{Law+2023} for drizzled MIRI-MRS data at the longest wavelength considered (\hfivea, $\lambda\approx7.5$~\micron). The smoothed cubes were then re-gridded to a common velocity axis, based on the \hfivea emission line.

\subsection{Ancillary ALMA data\label{sec:alma_data}}

We compare the JWST data with ALMA observations in multiple bands in order to compare the structure of the observed JWST emission with unobscured ionized gas emission as well as the cool molecular gas and dust. Datasets include observations at Band 3 (project code 2011.1.00172.S and 2012.1.00108.S, PI: A.~D. Bolatto; project code 2017.1.00895.S, PI: E.~A.~C. Mills), Band 6 (project code 2013.1.00191.S, PI: A.~D. Bolatto), Band 7 (project code 2015.1.00274.S, PI: A.~D. Bolatto), and Band 9 (project code 2018.1.00294.S, PI: A.~D. Bolatto). We focus especially on multiple transitions of CO to compare the distribution of the cool molecular component with the warm molecular gas and ionized gas traced by the MIRI-MRS data.

\subsubsection{Band 3\label{sec:band_3}}

Cycle 0 Band 3 ($\nu \approx 100$~GHz) observations of the central $\sim 1\arcmin$ (1~kpc) of NGC~253 (project code: 2011.1.00172.S) were carried out with a 16-antenna array in two frequency setups, each covering 8~GHz of instantaneous bandwidth. The first setup spanned 85.6--89.6~GHz and 97.4--101.4~GHz, observed as a 3-point mosaic along the major axis in the extended configuration at a resolution of $\sim 2\arcsec$ (35~pc). The second setup covered 99.8--101.3~GHz and 111.8--115.7~GHz using a 7-point mosaic in the compact configuration at a resolution of $\sim 4\arcsec$ (70~pc) over a field of view of $\sim 15\arcmin$ (1.5~kpc). Cycle 1 Band 3 observations (project code 2012.1.00108.S) were obtained in the extended configuration with 36 antennas and a total on-source integration time of $\sim 2.5$ hours, achieving a resolution of $\sim 2\arcsec$ (35~pc) in \COonezero and associated dense gas tracers. For both cycles, the interferometric \COonezero visibilities were combined with single-dish observations from Mopra (PI: J. Ott) to recover emission on all spatial scales. The combined dataset has a beam of 1\farcs85 $\times$ 1\farcs32 (31.5~pc $\times$ 22.4~pc), a spectral resolution of 5~\kms, and an rms noise of $\sim 2.5$~\mjyb per channel. In this paper, we focus on the \COonezero line at 115.271~GHz. More details on these data can be found in \citet{Bolatto+2013}, \citet{Leroy+2015}, and \citet{Meier+2015}.

Cycle 5 observations consist of a single pointing toward the nucleus of NGC~253 (project code 2017.1.00895.S), covering a $\sim 1\arcmin$ (1~kpc) field of view in three extended configurations (C43-6, C43-7, C43-8; with baselines 15--8500m). The imaged data have beam sizes of 0\farcs21 $\times$ 0\farcs15 (3.6~pc $\times$ 2.5~pc) and are sensitive to size scales up to $\sim$ 2\farcs6 (44~pc) at 100~GHz. Four 1.875~GHz wide sub-bands were centered at 86.63, 88.48, 98.52, and 100.38~GHz, with a spectral resolution of 0.977~MHz (2.8--3.4~\kms). Visibilities from these observations were pipeline calibrated using the Common Astronomy Software Application \citep[CASA;][]{McMullin2007}) version 5.4.0. In this paper, we focus on the continuum image, with an effective central frequency of 94~GHz. More details of the data calibration and imaging can be found in \citet{Mills+2021}.

\subsubsection{Band 6\label{sec:band_6}}

Cycle 2 ALMA Band 6 ($\nu \approx 230$~GHz) observations of NGC~253 targeted the \COtwone line at 230.538~GHz using a 14-point 12m mosaic with Nyquist sampling, supplemented by Atacama Compact Array (ACA) and Total Power (TP) observations to recover emission on all spatial scales. The 12m observations used 36 antennas and were carried out on 2014 December 28, with ACA observations on 2014 June 28 and TP observations between 2015 May and August. The center of the observed region is offset from the photometric center of NGC~253 in order to cover the full extent of the southern molecular outflow. Calibration of the \COtwone data used scripts provided by ALMA staff for the ACA data and the ALMA pipeline for the 12m data, with initial calibration and imaging performed in CASA version 4.2.2. The 12m and ACA data were combined and imaged using Briggs weighting (\texttt{robust=0.5}), then combined with the TP data using \texttt{FEATHER} in CASA version 4.7.2. The original synthesized beam is 1\farcs7 $\times$ 1\farcs0 (PA~$= 80\fdg9$) with an rms of 2.0~\mjyb per 5~\kms~channel. The final cubes are smoothed to 1\farcs9 $\times$ 1\farcs4 (32~pc $\times$ 24~pc) at PA~$= 78$\degree to match the \COonezero resolution, yielding an rms of 2.2~\mjyb per channel. In this paper, we focus on the \COtwone line. More details on these data can be found in \citet{Zschaechner+2018}.

\subsubsection{Band 7\label{sec:band_7}}

Cycle 3 ALMA Band 7 ($\nu \approx 350$~GHz) observations targeted the central $\sim 750$~pc of NGC~253 with a 4-point 12m mosaic (compact and extended configurations) and a 5-point ACA mosaic. The baseline ranges are 8.9--49.0m for the ACA, 15.1--783.5m for the 12m compact array, and 15.1--1813.1m for the 12m extended array. Additional TP observations recover large scale emission. The imaged data have beam sizes of $\sim$ 0\farcs175 (3~pc) and are sensitive to size scales up to $\sim$ 19\farcs6 (320~pc). The spectral setup spans 342.0 to 357.7~GHz with a 976.6~kHz channel width (0.8~\kms). The visibilities of the 12m data are calibrated using the ALMA Cycle 3 pipeline in CASA 4.6.0, and the other datasets are calibrated in CASA 4.7.2 and the Cycle 4 pipeline. In this paper, we focus on the \COthreetwo line at 345.79599 GHz. More information on these data can be found in \citet{Leroy+2018} and \citet{Krieger+2019}.

\subsubsection{Band 9\label{sec:band_9}}

Cycle 6 ALMA Band 9 ($\nu \approx 690$~GHz) observations cover the central $16\arcsec \times 13\arcsec$ region using a 3-pointing 12m mosaic in the C43-5 configuration. The imaged data have beam sizes of $\sim$ 0\farcs11 (1.9~pc) and a maximum recoverable scale of $\sim$ 1\farcs47 (25~pc). The spectral range spans 688.19 to 695.45~GHz. Visibilities were combined and pipeline calibrated using the CASA pipeline version 2020.1.0.40. Before imaging, spectral windows that may contain the \COsixfive line were flagged \citep[assuming a galaxy recessional velocity of 243~\kms;][]{Koribalski+2004}. Imaging was performed in CASA 5.4.1 using \texttt{tclean} with 0\farcs02 cell size, \texttt{mfs} mode, Hogbom deconvolver, and Briggs weighting (\texttt{robust=0.5}). Masks were auto-generated (\texttt{auto-multithresh}) and cleaned down to 0.3~\mjyb, yielding a residual rms of 0.26~\mjyb (54~mK). In this paper, we focus on the \COsixfive line at 691.47308 GHz. More information on this dataset can be found in K.~Donaghue et al.\ in prep.

\subsection{Archival HST Data and Alignment\label{sec:HST_data}}

In comparing the JWST data to existing HST-NICMOS data, a spatial offset was noted between the structures visible in hydrogen recombination lines in both data sets. We cross-correlated the NICMOS H3$\alpha$ (i.e., Paschen-$\alpha$, $\lambda=1.87$~\micron; HST Program 7218, PI: M.~J. Reike) line map with the \hfivea (i.e., Pfund-$\alpha$, $\lambda=7.46$~\micron) line map from MIRI-MRS and found the HST astrometry is offset relative to the JWST data by $\Delta \alpha, \Delta \delta = +$ 0\farcs78, $-$ 0\farcs2.

\section{Results and Analysis\label{sec:results_analysis}}

\begin{figure*}[htbp]
    \centering
    \includegraphics[width=0.51\textwidth]{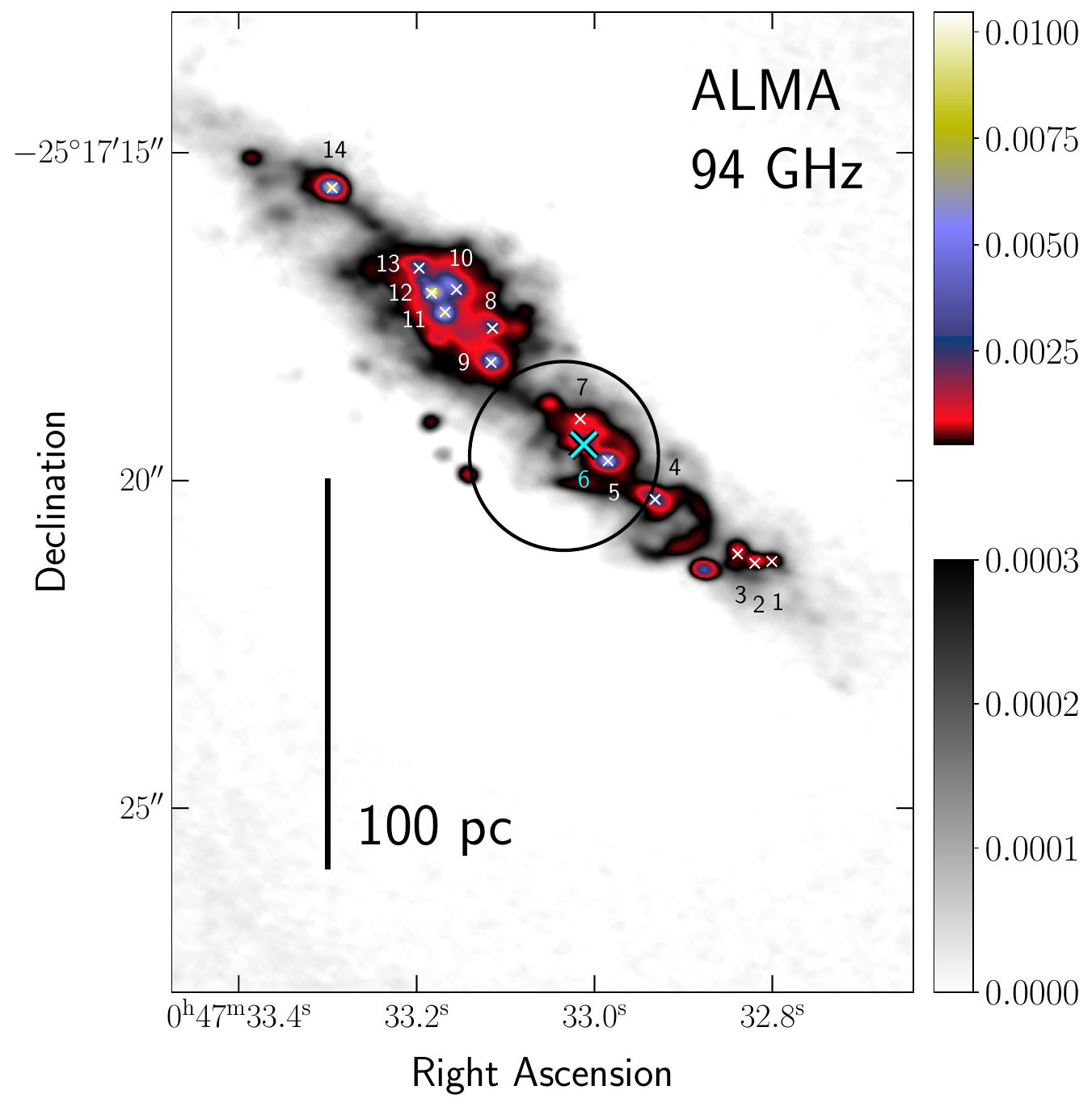}\includegraphics[width=0.49\textwidth]{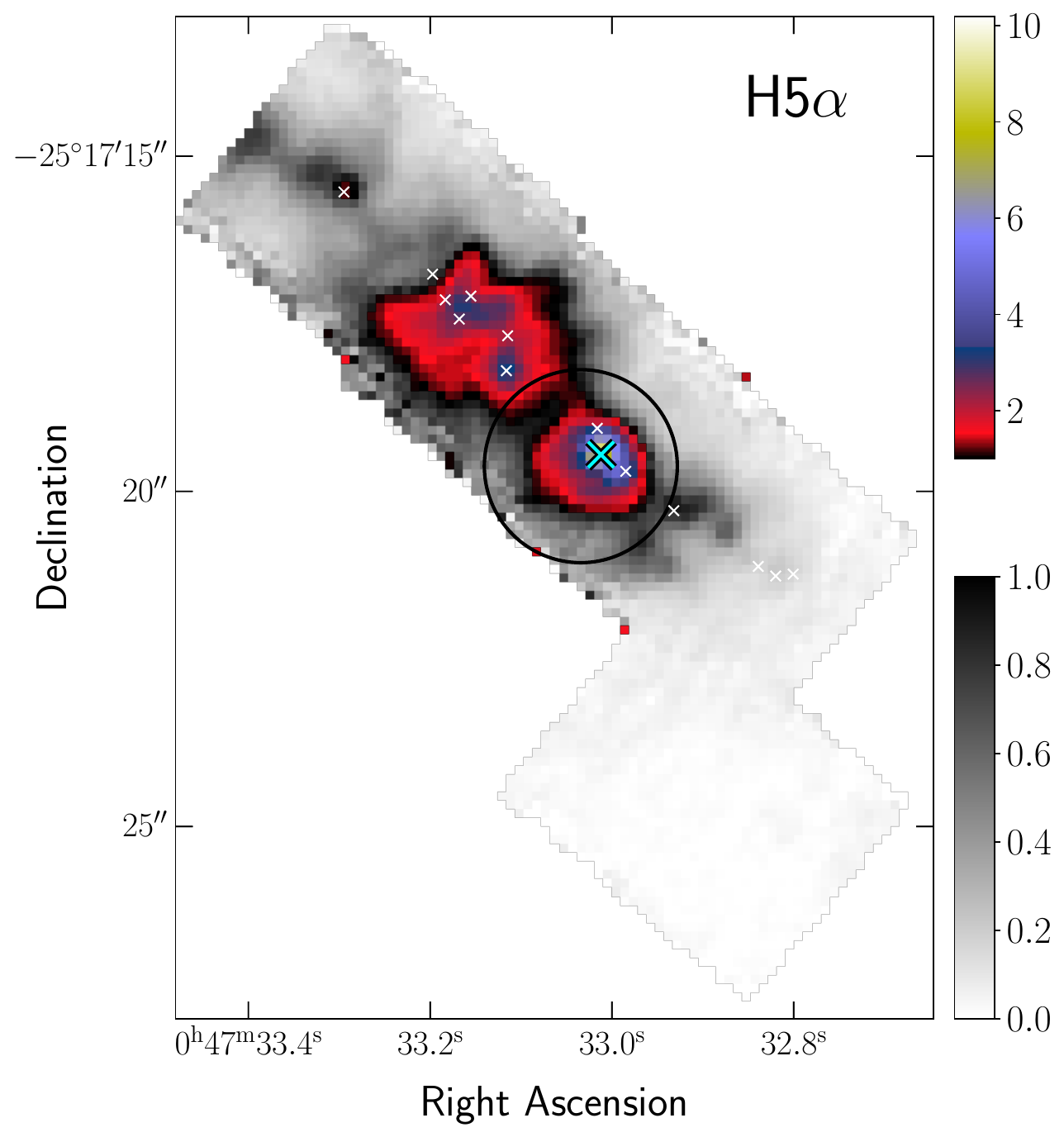}\\
    \includegraphics[width=0.51\textwidth]{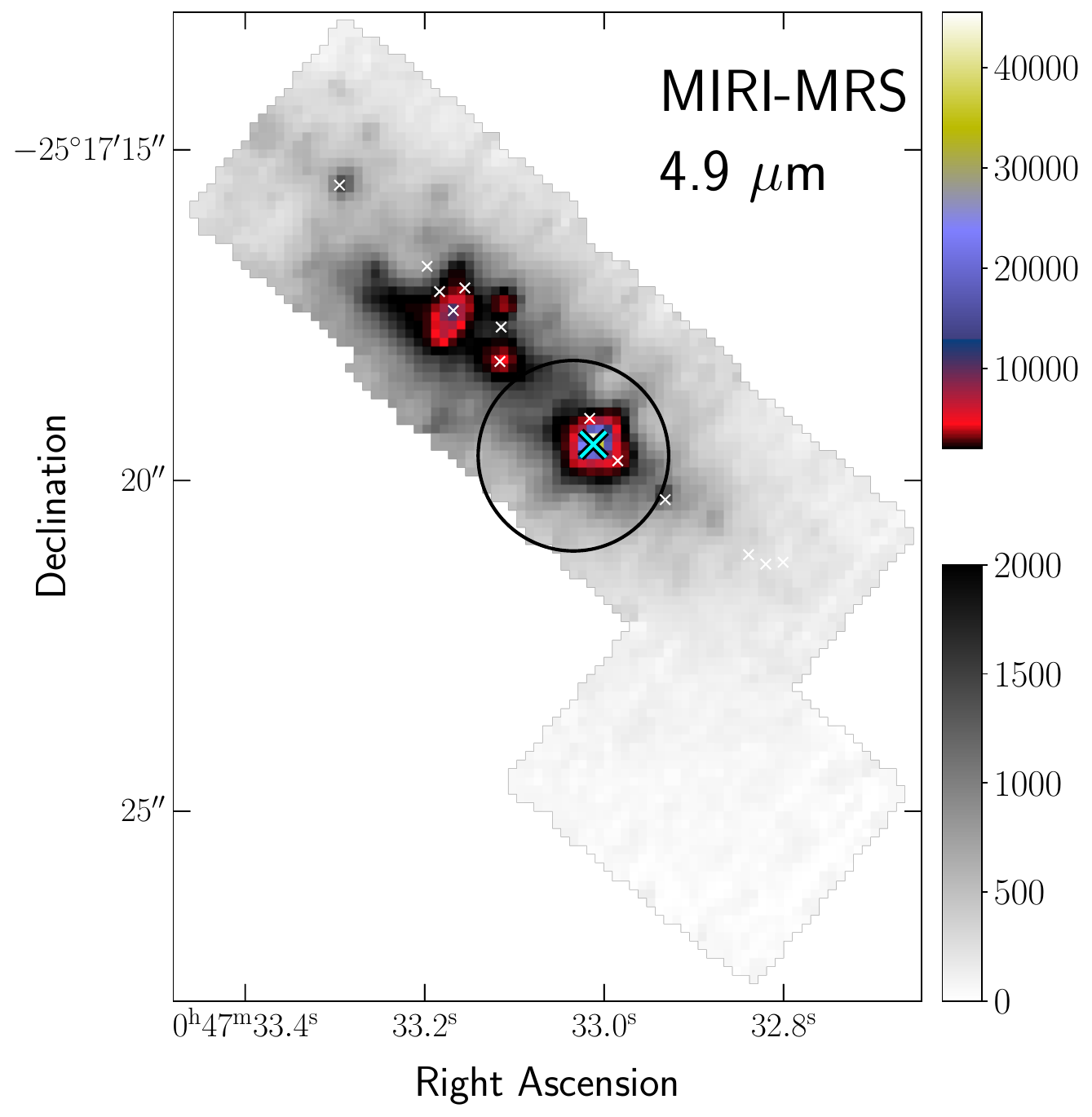}\includegraphics[width=0.49\textwidth]{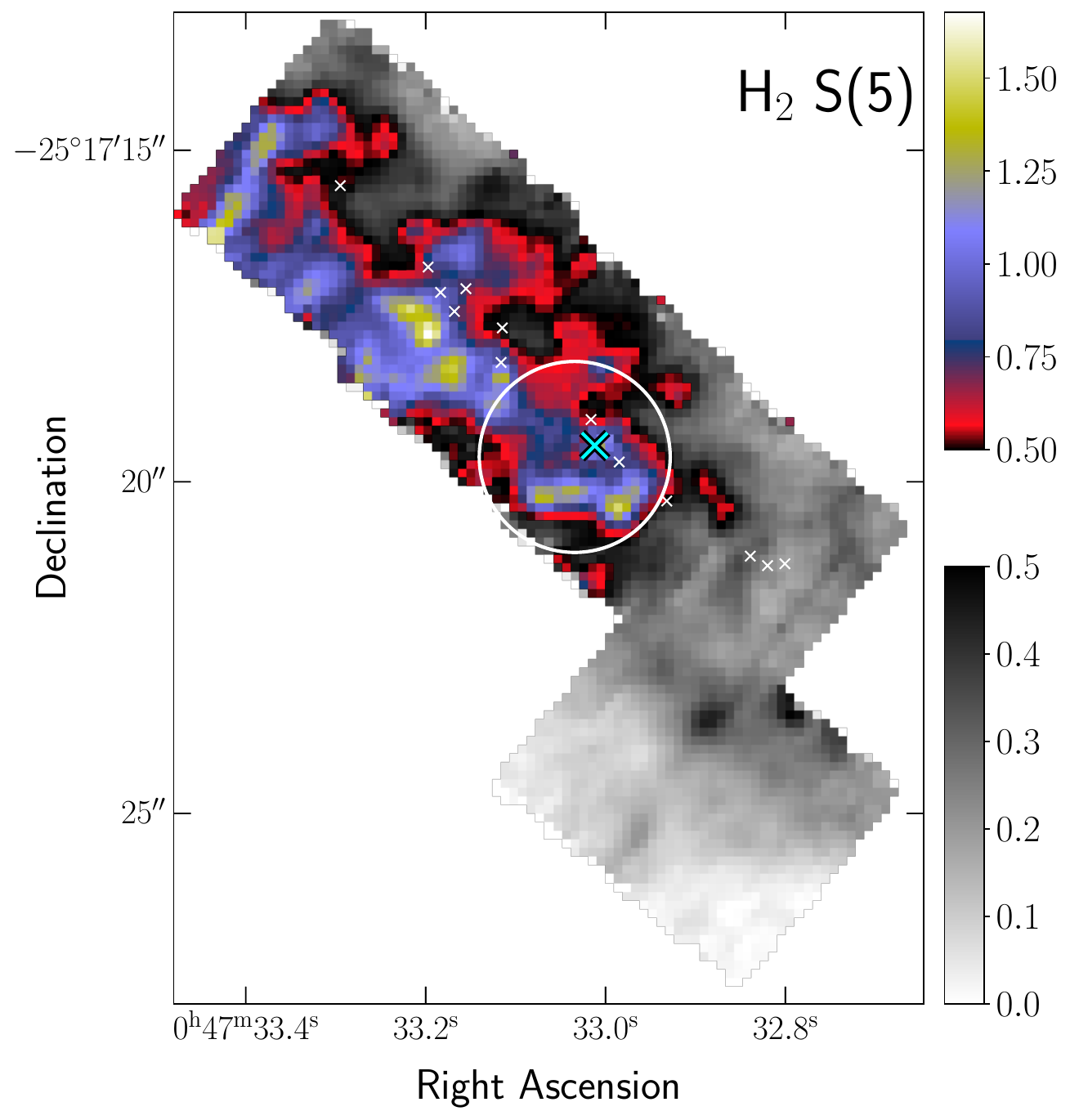}
    \caption{\textbf{Top Left:} ALMA Band 3 continuum image from \citet{Mills+2021}, tracing a mix of free-free continuum emission from ionized gas, dust thermal emission from compact gas clouds, and synchrotron emission from supernova remnants. \textbf{Top Right:} Continuum-subtracted and integrated emission from the \hfivea line tracing ionized gas with units of $10^6$~\Mjysr~\kms. \textbf{Bottom Left:} MIRI-MRS continuum image of the central starburst of NGC~253 at 4.9~$\micron$, primarily showing stellar emission, with units in \Mjysr. \textbf{Bottom Right:} Continuum-subtracted and integrated emission from the \htwoSfive line tracing warm molecular gas with units of $10^6$~\Mjysr~\kms. Cluster positions from \citet{Mills+2021} for the 14 clusters identified from \citet{Leroy+2018} are plotted as white x-shaped points and numbered in the top left panel. The position of SSC~6 is shown as an oversized cyan x. The circle indicates the region that is further extracted for analysis in this work.}
    \label{fig:bubble_identification}
\end{figure*}

\subsection{Identification of a Bubble-like Structure\label{sec:bubble_id}}

In the \hfivea map (top right panel of Figure~\ref{fig:bubble_identification}), we identify by-eye an extended structure which appears as a bright peak surrounded by an extended envelope of emission that is also spatially coincident with a shell-like structure in the ALMA Band 3 continuum data (top left panel of Figure~\ref{fig:bubble_identification}). The extended emission encompasses three SSCs (5, 6, and 7) previously identified with ALMA \citep{Leroy+2018} and is roughly centered on the position of SSC~6. In the MIRI–MRS continuum image of stellar emission (bottom left panel of Figure~\ref{fig:bubble_identification}), the brightest point source, which we associate with the cluster previously identified at near-infrared wavelengths by \citet{Watson+1996} and \citet{Kornei+2009}, also corresponds to ALMA SSC~6. We note that this spatial association differs from the interpretation presented by \citet{Leroy+2018} and \citet{Levy+2021}, in which they associate the infrared cluster with ALMA SSC~5. The implications of this revised association will be discussed further by R.~C.~Levy et al.\ in prep. 

\begin{figure*}[htbp]
    \centering
    \includegraphics[width=\textwidth]{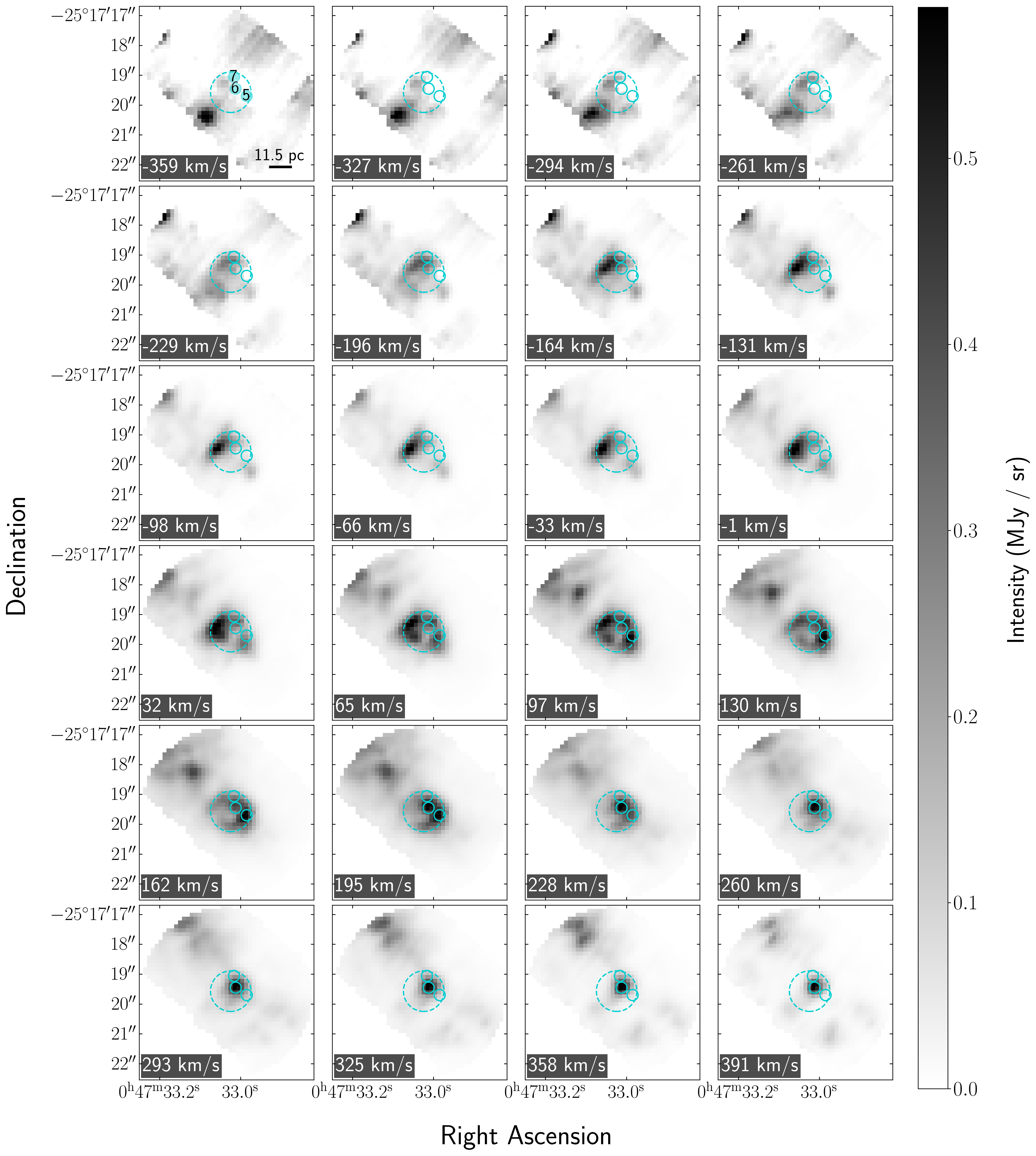}
    \caption{\arii (7.0~$\micron$) emission line channel maps with units of \Mjysr and velocity range ($-359$, $+391$)~\kms. The larger dashed cyan circle represents the approximate extent of the bubble emission. A scale bar of 11.5~pc (bubble radius) is given in the first panel. The smaller solid cyan circles are the central positions of SSC~5, 6, and 7, numbered in the first panel. This extracted region is defined in Figure~\ref{fig:bubble_identification}.}
    \label{fig:ArII_channel_maps}
\end{figure*}

We construct velocity channel maps for each of the molecular and ionized tracers listed in Table~\ref{tab:line_properties}. Among these, \arii as the brightest line provides the clearest view of the morphology, revealing coherent emission across a broad velocity range of $-250$~\kms to $+350$~\kms in Figure~\ref{fig:ArII_channel_maps}. The emission appears consistent with a shell or bubble, with the brightest emission primarily concentrated along the bubble's edge, and especially prominent at velocities from $\sim$30--190~\kms. The bubble appears slightly flattened, with the major axis roughly aligned with the plane of the galaxy.

\begin{figure*}[htbp]
    \centering
    \includegraphics[width=\linewidth]{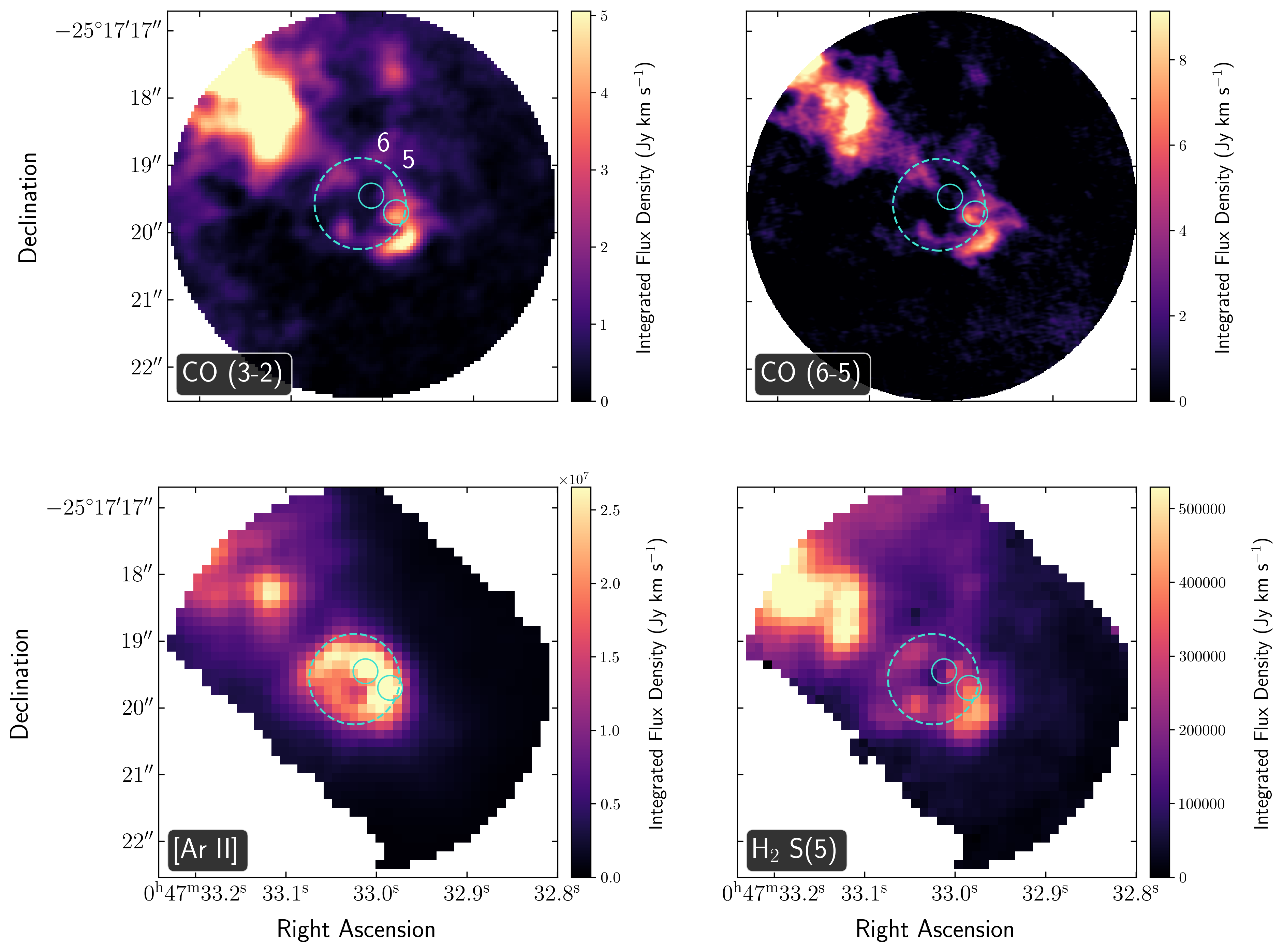}
    \caption{Restricted-velocity (25--198~\kms) moment 0 maps for selected ionized and molecular tracers in units of \jykms. The velocity range was chosen to isolate the bubble-associated emission from other bright features present in the spectrum. All maps are labeled in the bottom left corner. The location and size of the bubble is shown with the dashed cyan circle (see Table~\ref{tab:bubble_properties}. The solid cyan circles are the central positions of SSC~5 and 6, numbered in the first panel.}
    \label{fig:moment0_map}
\end{figure*}

Integrating over a velocity range between 25--198~\kms reveals a consistent shell or bubble-like morphology in both molecular and ionized gas (Figure~\ref{fig:moment0_map}). \htwo and CO emission is concentrated along the rim of the bubble, with the brightest emission appearing in the southwest region, forming a shell of swept-up material. This differs from the ionized \arii emission which is more concentrated toward the location of SSC~6, forming an ionized bubble of emission. We therefore refer to ionized emission as the ``bubble" and molecular emission as the (swept-up) ``shell". The limb-brightening of the bubble is not completely uniform around the periphery, potentially reflecting variations in density of swept-up material or in the ionizing flux, given the apparent northward offset of SSC~6 from the bubble's geometric center. Alternatively, the asymmetry may indicate breakout along its southeast edge (Section~\ref{sec:coupling}).

\subsection{Forward Modeling of the Bubble\label{sec:forward_modeling}}

To put some quantitative limits on the physical and kinematic properties of the bubble, we construct a simple three-dimensional model and compare it directly to our observations. For simplicity, the bubble is represented as a spherical shell of uniform emissivity (even though it appears slightly elongated along the plane of the galaxy), parameterized by its radius $r_0$, defined as the shell midpoint such that the shell spans $r_0 - \Delta r / 2$ to $r_0 + \Delta r / 2$, its shell thickness $\Delta r$, fill fraction $f$ (the relative emissivity of gas interior to the shell), and center offset $(\delta x, \delta y)$ from the by-eye estimated bubble center at ($\alpha$, $\delta$) = (11.8876126\degree, $-$25.2887904\degree), used as a fixed reference point for the fit. We discretize the model onto a Cartesian grid with a fine pixel scale of 0.1~pc and, for kinematics, assign a radial expansion velocity of $v_{\rm exp}$ to each shell voxel, producing a model cube in $(x, y, v_{\rm LOS})$ space. The cube is then convolved to match the spectral line spread function of the instrument and resampled to the observed spatial and spectral pixel scales (2.2~pc~pix$^{-1}$ and $\sim 34$~\kms~channel$^{-1}$, respectively) so that the synthetic observations are directly comparable to the data.

We fit the model in two sequential stages. First, we determine the morphological parameters ($r_0$, $\Delta r$, $f$, and center offset) by minimizing the root mean square (RMS) residual between the flux-normalized model moment 0 map and the observed moment 0 map. Second, using our best-fit morphology parameters, we determine the kinematic parameters ($v_{\rm exp}$ and $v_{\rm sys}$) by minimizing the RMS residual in position-velocity (PV) space. We adopt a grid-search approach throughout and estimate parameter uncertainties from the curvature of the RMS profile projected onto each parameter axis, using the FWHM of the likelihood curve (Section~\ref{sec:morphology_fit}, \ref{sec:kinematic_fit}). All best-fit values and uncertainties are reported in Table~\ref{tab:modeling_fit}. We find that the best-fit bubble center lies at an offset of $\sim$0\farcs20 (3.4~pc) from SSC~6, well within the bubble radius (11.5 $\pm$ 3.4~pc), supporting SSC~6 as a source powering the bubble's expansion.

\begin{table}[htbp]
\begin{threeparttable}
\caption{Forward modeling best-fit results.}
\label{tab:modeling_fit}
\small
\setlength{\tabcolsep}{4pt}
\begin{tabular}{l l}
\toprule
\textbf{Property} & \textbf{Best-Fit Value} \\
\midrule
Radius [pc] & 11.5 $\pm$ 3.4 \\
Thickness [pc] & 4.7 $\pm$ 2.9 \\
Interior Filling Fraction\textsuperscript{a} [\%] & 68$^{+32}_{-43}$ \\
Offset from SSC 6\textsuperscript{b} [arcsec] & 0.20 \\
Offset from SSC 6\textsuperscript{b} [pc] & 3.4 \\
$v_{\rm exp}$ [\kms] & 90 $\pm$ 44 \\
$v_{\rm sys}$ [\kms] & 231 $\pm$ 28 \\
\bottomrule
\end{tabular}
\begin{tablenotes}
\footnotesize
\item \textsuperscript{a} The interior emissivity fraction $f$ is the ratio of the interior volume emissivity to the shell emissivity.
\item \textsuperscript{b} Center offset is measured relative to SSC~6 (RA = 11.88755\arcdeg, Dec = $-$25.28873701\arcdeg).
\end{tablenotes}
\end{threeparttable}
\end{table}

\subsubsection{Morphological Fitting\label{sec:morphology_fit}}

Morphological fitting is performed only on our \arii moment 0 map, as the ionized gas most clearly traces the overall bubble structure and geometry. The resulting best-fit morphology is then adopted for subsequent PV fitting of the \arii ionized gas component.

The model moment 0 map is flux-normalized and compared to the observed moment map (masked at 30\% peak intensity), giving us the RMS of the residuals. We also mask out the brightest 4$\times$4 pixels surrounding SSC~6 in order to better isolate the bubble emission. Parameter uncertainties are estimated from the width of the marginal RMS profiles. We convert the relative RMS degradation into a likelihood weighting, $\mathcal{L} \propto \exp(-\Delta \mathrm{RMS} / \mathrm{RMS}_{\rm min})$ and use the FWHM of this curve as an empirical measure of the parameter constraint.

\begin{figure*}[htbp]
    \centering
    \includegraphics[width=\linewidth]{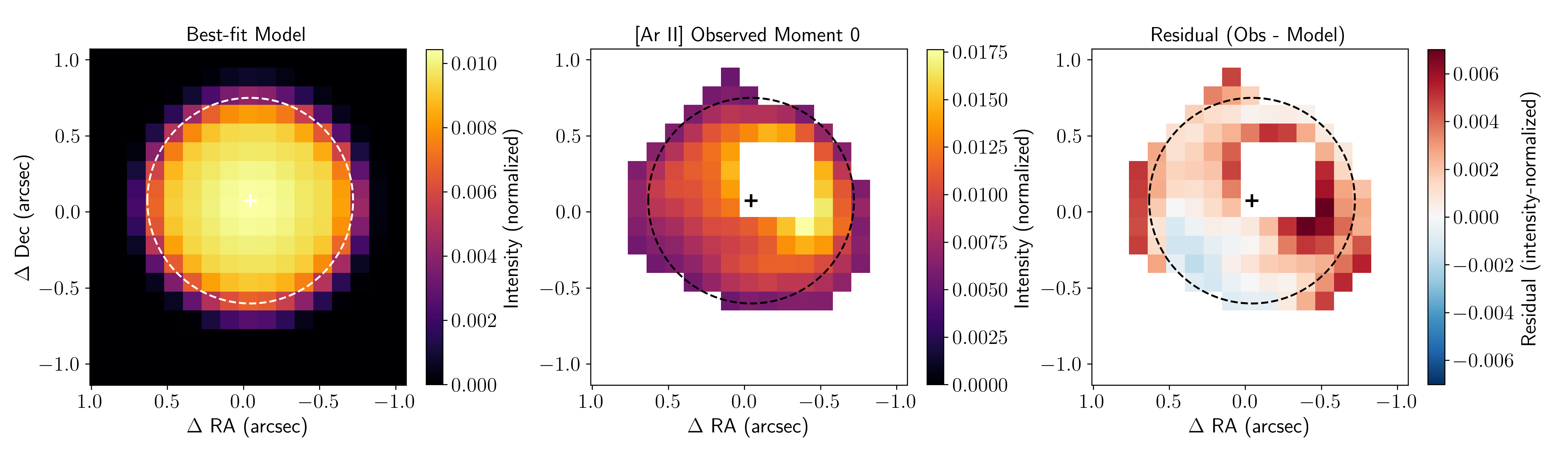}
    \vspace{0.3cm} \\
    \includegraphics[width=\linewidth]{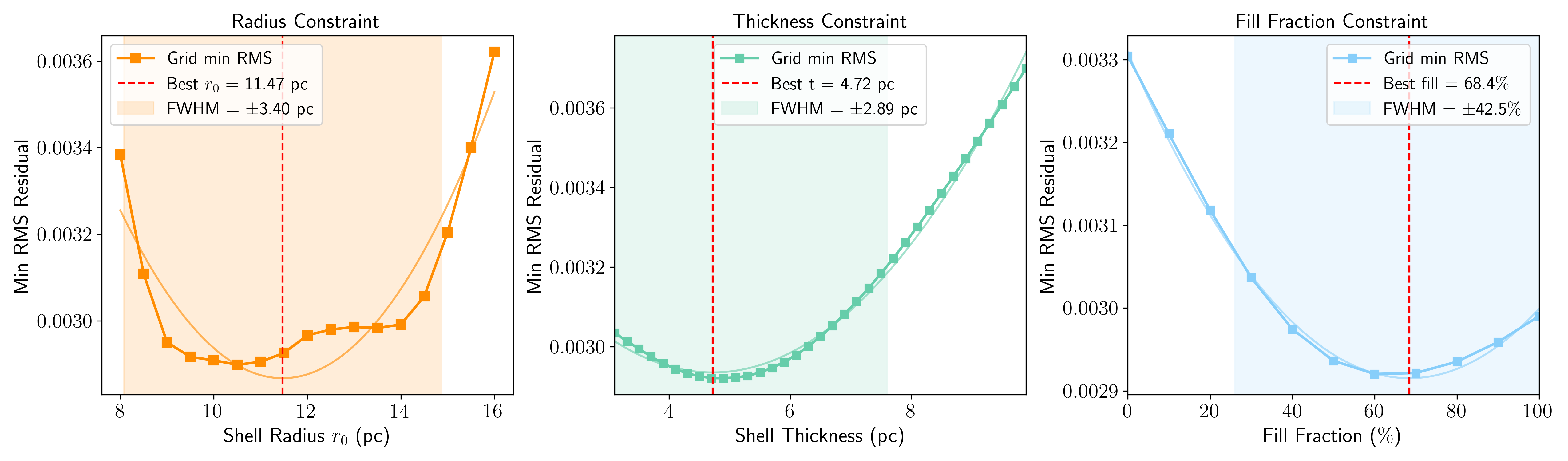}
    \caption{\textbf{Top Row:} Forward modeling morphological fit to the \arii moment 0 map. The Best-fit model moment 0 map (\textit{left}), observed \arii moment 0 map (\textit{center}), and pixel-by-pixel residual map (\textit{right}) are all flux-normalized. The dashed circle marks the best-fit shell radius $r_0$, and the cross marks the best-fit center offset. The residual is shown on a scale where red (blue) indicates regions where the observation exceeds (falls below) the model. \textbf{Bottom Row:} Marginal RMS residual profiles for the shell radius $r_0$ (\textit{left}), shell thickness $\Delta r$ (\textit{center}), and the fill fraction $f$ (\textit{right}) with the parabolic fit (solid curve), best-fit value (red-dashed vertical line), and $1\sigma$ uncertainty range (shaded) indicated on each panel.}
    \label{fig:morphology_fit}
\end{figure*}

Figure~\ref{fig:morphology_fit} shows the best-fit model and residual map and the marginal RMS profiles for the radius, thickness, and filling factor constraints, which are summarized in Table~\ref{tab:modeling_fit}.

The best-fit radius of $r_0 = 11.5 \pm 3.4$~pc and shell thickness $\Delta r = 4.7 \pm 2.9$~pc indicate a relatively thick shell, with $\Delta r/r_0 \sim 0.4$, suggesting that the observed structure is not well described by a geometrically thin shell. This broad shell may reflect the intrinsic structure of the bubble as well as broadening associated with turbulence, density inhomogeneities, or other pressure-support effects in the surrounding ISM. Our radius is consistent with \cite{Gunthardt+2015}, who measure a comparable radius (11--16~pc) using \neii\ line emission. The best-fit interior filling fraction ($f = 68^{+32}_{-43}\%$) indicates that the observed morphology favors substantial emission from within the shell but does not strongly constrain the relative interior emissivity. In particular, the data do not rule out either a partially filled or nearly fully-filled bubble.

\subsubsection{Kinematic Fitting\label{sec:kinematic_fit}}

\begin{figure*}[htbp]
    \centering
    \includegraphics[width=\linewidth]{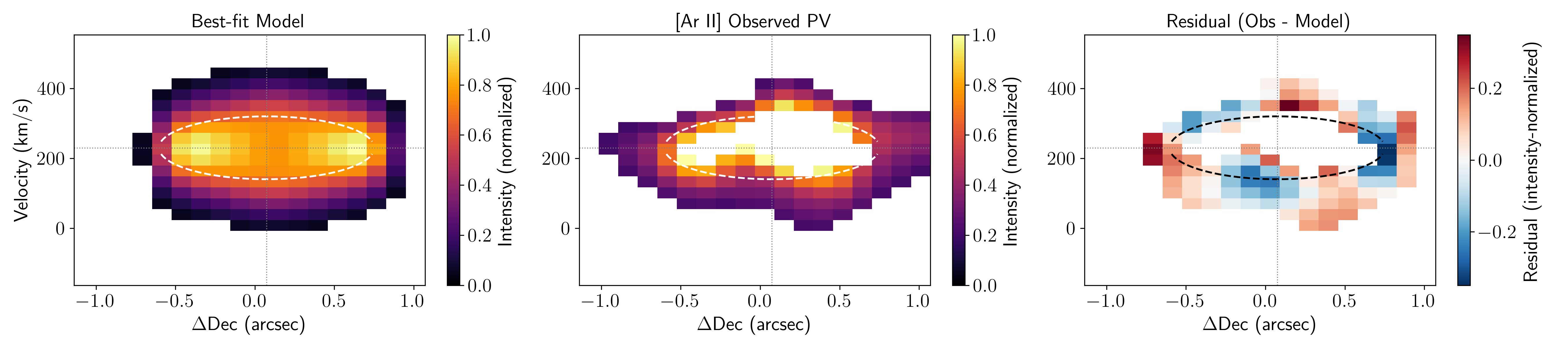}
    \caption{Forward modeling kinematic fit to the \arii PV diagram. The Best-fit model PV diagram (\textit{left}), observed \arii PV diagram (\textit{center}), and pixel-by-pixel residual map (\textit{right}) are all flux-normalized. The best-fit dashed ellipse, with semi-axes $r_0$ and $v_{\rm exp}$ centered on $v_{\rm sys}$, is over-plotted on each panel. The residual is shown on a scale where red (blue) indicates regions where the observations exceeds (falls below) the model.}
    \label{fig:kinematic_fit}
\end{figure*}

\begin{figure}[htbp]
    \centering
    \includegraphics[width=\linewidth]{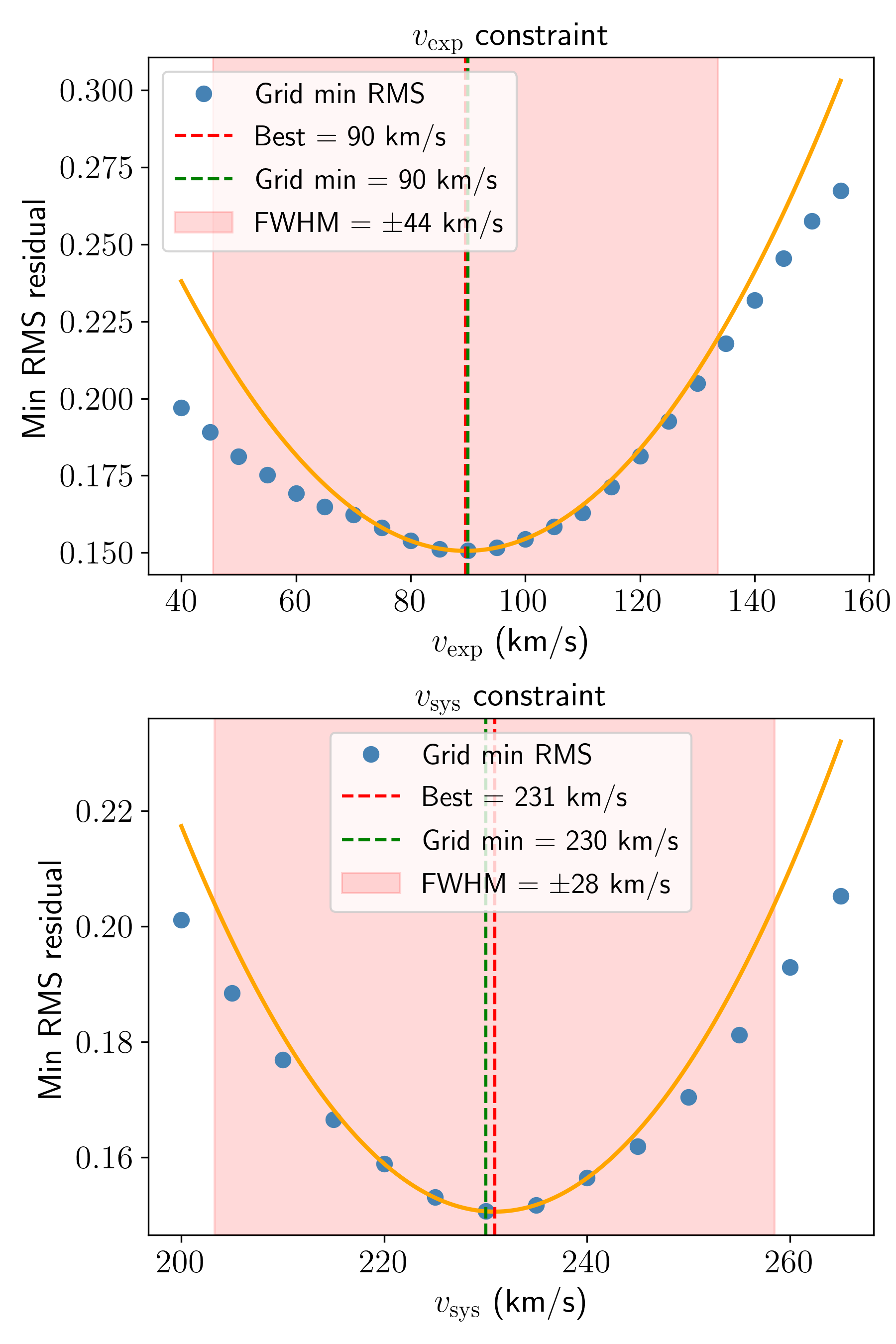}
    \caption{Marginal RMS residual profiles for the expansion velocity $v_{\rm exp}$ (\textit{top}) and systemic velocity (\textit{bottom}), with the parabolic fit (solid yellow curve), best-fit value (red-dashed vertical line), grid minimum (green-dashed vertical line), and $1\sigma$ uncertainty range (shaded) indicated on each panel.}
    \label{fig:kinematic_constraint}
\end{figure}

With the morphological parameters fixed at their best-fit values, we search a grid of expansion velocity $v_{\rm exp}$ (40--160~\kms) and systemic (shell) velocity $v_{\rm sys}$ (200--270~\kms) by comparing model and observed PV diagrams extracted along the north-south direction through the bubble center. An expanding spherical shell produces a characteristic ellipse in PV space with spatial semi-axis $r_0$ and velocity semi-axis $v_{\rm exp}$. We generate a fresh model cube at each trial $v_{\rm exp}$ and interpolate it onto the observed $(y,v)$ grid for each $v_{\rm sys}$, which is implemented as a shift of the model velocity axis. Contaminating emission from the central cluster is masked within a $\pm 50$~\kms window centered on $v_{\rm sys}$ over the cluster's spatial extent, and a low-level brightness threshold is applied before comparison. The RMS residual between the normalized model and observed PV maps is computed over all valid spaxels. We verified that our results are insensitive to surface brightness weighting and masking threshold (2--20\% of peak intensity), with best-fit values shifting by only a few \kms. Parameter uncertainties are estimated from the marginal RMS profiles using the same FWHM-likelihood method adopted for the morphological fit. Figure~\ref{fig:kinematic_fit} shows the best-fit model PV diagram, observed PV diagram, and residuals for \arii, while Figure~\ref{fig:kinematic_constraint} presents the marginal RMS profiles for $v_{\rm exp}$ and $v_{\rm sys}$. All kinematic best-fit values are reported in Table~\ref{tab:modeling_fit}.

\begin{figure}[htbp]
    \centering
    \includegraphics[width=\linewidth, trim={8cm 0.5cm 8cm 0.5cm}, clip]{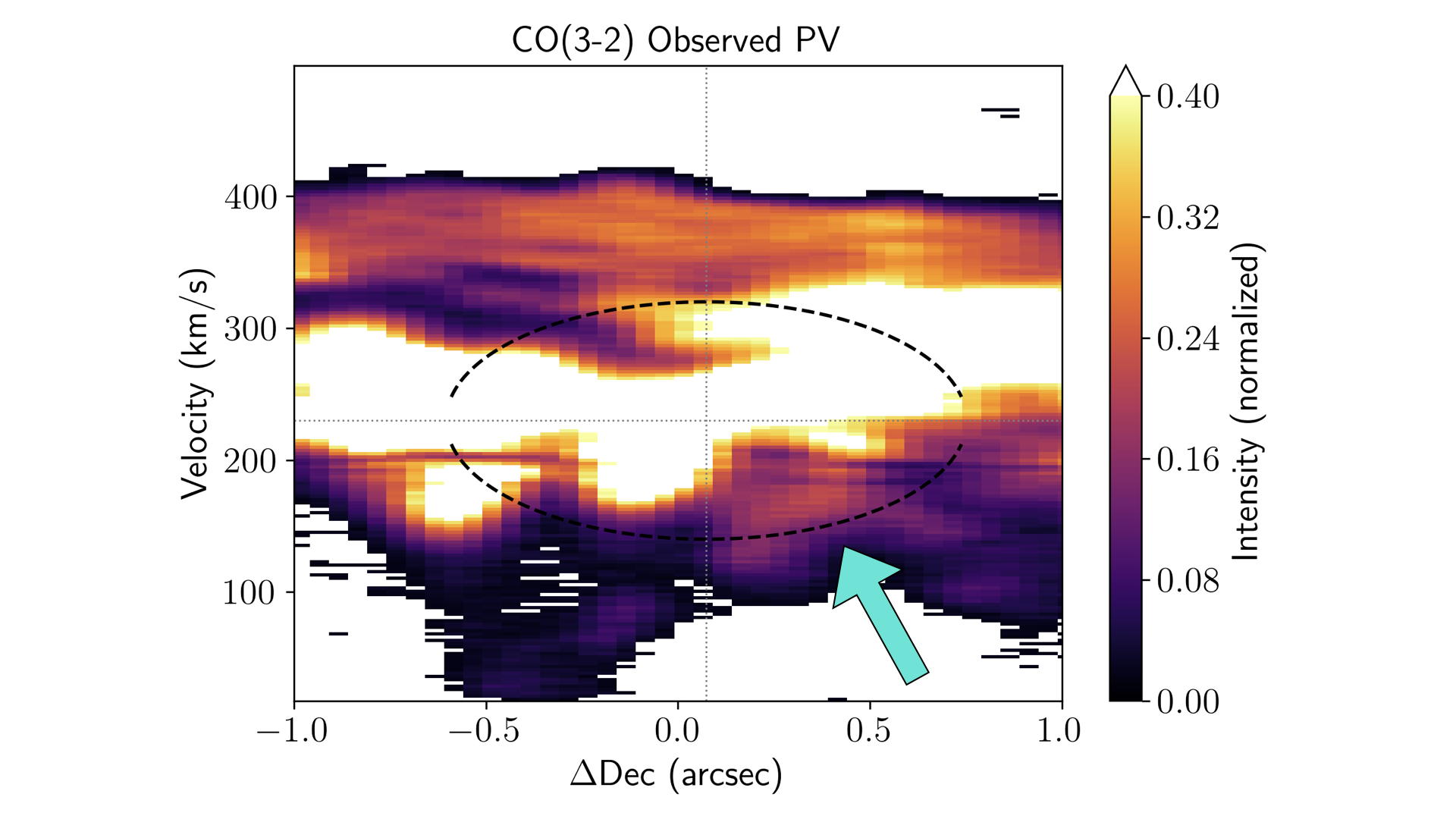}
    \caption{Observed \COthreetwo PV diagram extracted along the north-south direction through the bubble center, with the best-fit \arii ellipse (dashed) overlaid for reference. The cyan arrow marks faint emission at lower velocities ($\sim 100$~\kms) offset to the south, which we associate with the swept-up shell. A bright, spectrally broad component dominates at velocities $\gtrsim 200$~\kms, likely unrelated foreground molecular emission. For this reason, we saturate (in white) the brightest emission in the PV diagram.}
    \label{fig:CO_model_overlay}
\end{figure}

Our simple model has difficulty fitting to the observed PV data, which we attribute to several factors: the difficulty in fully masking out the bright SSC~6 source, the non-uniform limb-brightening of the bubble, related to the likely breakout at its southeastern edge (Section~\ref{sec:coupling}), where we see high-velocity emission ($v < 100$ \kms) in Figures~\ref{fig:ArII_channel_maps} and \ref{fig:CO_model_overlay} that is not well captured by the model. The best-fit expansion velocity of $v_{\rm exp} = 90 \pm 44$~\kms is consistent with energetic stellar feedback from a young SSC, and the systemic velocity of $v_{\rm sys} = 231 \pm 28$~\kms is consistent with the known kinematics of the gas immediately surrounding SSC~6 \citep{Leroy+2018, Mills+2021, Levy+2021, Levy+2022}, confirming the bubble's physical association with the cluster.

Figure~\ref{fig:CO_model_overlay} shows the \COthreetwo PV diagram with the best-fit ionized gas ellipse overlaid. The molecular emission is dominated by a bright, broad component above $\sim 200$~\kms that likely corresponds to gas orbits on a ring in the inner 200~pc of the galaxy \citep{Levy+2022}, and which prevents reliable kinematic modeling. Faint emission at $\sim 100$~\kms (cyan arrow) offset to the south is associated with the molecular shell, consistent with the expanding shell seen in CO in Figure~\ref{fig:moment0_map}. The spatial and velocity extent of this feature is consistent with the best-fit ellipse from the ionized gas. The absence of a clear CO counterpart to the full ionized shell reinforces the interpretation that the ionized gas traces the directly driven inner shell, while CO traces only the densest portions of the surrounding swept-up molecular medium.

\subsection{Non-LTE modeling of CO \label{sec:radex_modeling}}

To constrain the physical conditions of the molecular gas associated with the bubble, we perform a non-local thermodynamic equilibrium (non-LTE) excitation analysis using the radiative transfer code \texttt{RADEX} \citep{van_der_Tak+2007}. \texttt{RADEX} solves the statistical equilibrium equations under the escape probability approximation and allows us to estimate the CO column density by comparison with line intensities of multiple observed transitions.

\begin{figure}
    \centering
    \includegraphics[width=\linewidth]{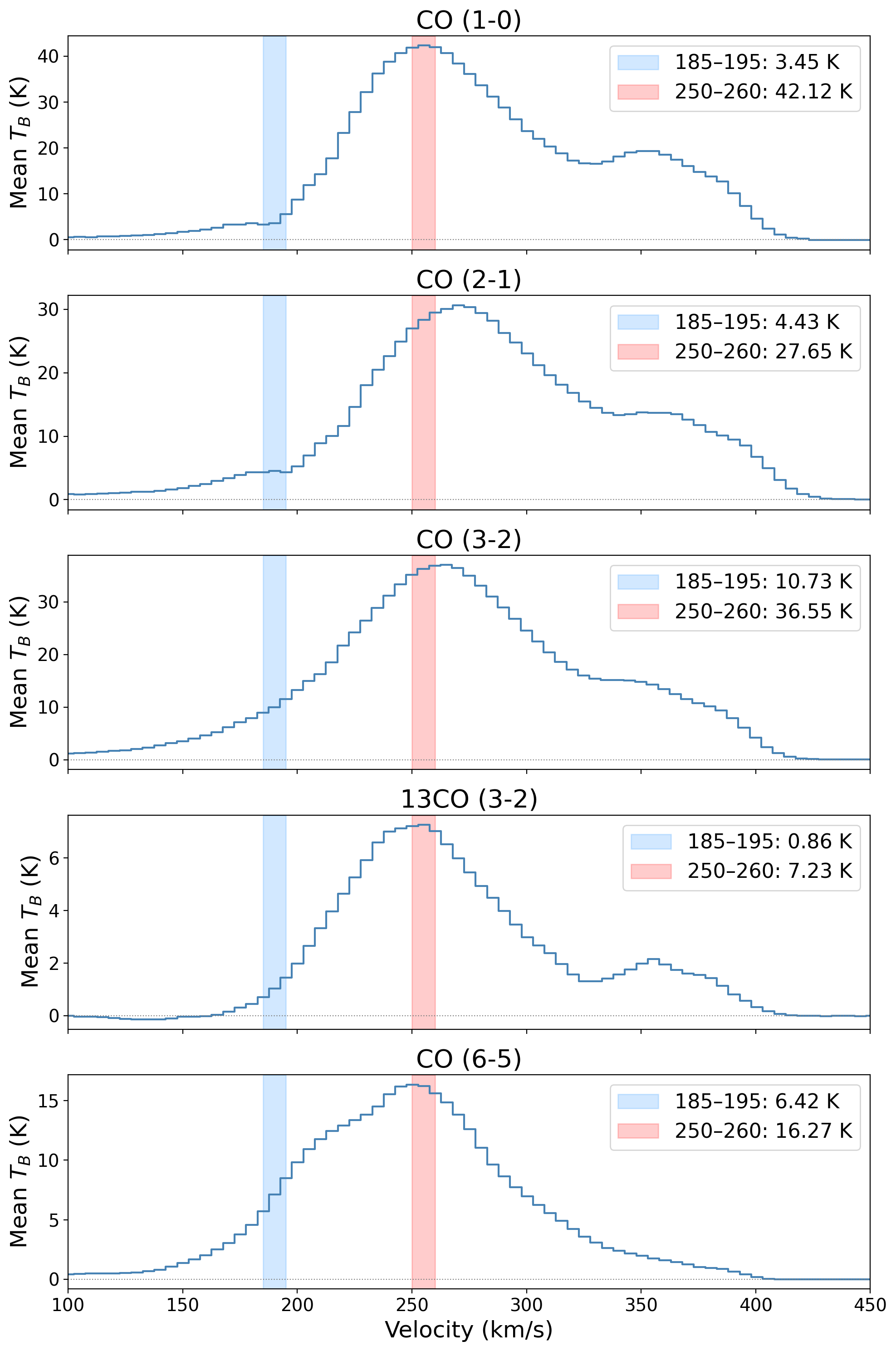}
    \caption{Representative spectra for each CO transition within the bubble aperture, constructed by taking the mean brightness temperature at each velocity over all pixels within the aperture. The red shaded regions (250--260~\kms) indicate the velocity interval used to extract the representative line intensities for the primary non-LTE analysis, corresponding to the peak of the molecular emission. The blue shaded regions (185--195~\kms) indicate the velocity interval used for the second non-LTE analysis, corresponding to the lower-velocity emission associated with the expanding shell (Figure~\ref{fig:moment0_map}). The secondary component is most clearly distinguished in the \COonezero and \COtwone spectra.}
    \label{fig:radex_spectra}
\end{figure}

A three-dimensional parameter grid was generated across a wide range of molecular gas conditions \citep[that bracket known conditions in the overall ISM; i.e.,][]{Ott+2005, Mangum+2019, Tanaka+2024}. The kinetic temperature was allowed to vary from 30 to 300~K in 40 linearly spaced steps, the molecular hydrogen volume density was varied from $10^2$ to $10^7$~\cm in 40 logarithmically spaced steps, and the CO column density ranged from $10^{15}$ to $10^{20}$~cm$^{-2}$ in 20 logarithmically spaced steps. We additionally generated a grid of filling factors ranging from 0.01 to 1.0, sampled using 20 linearly spaced values. These were used to account for potential impacts of beam dilution of the observations. Throughout the modeling, we: (1) assumed a cosmic microwave background radiation field of 2.73~K and (2) adopted a characteristic line width of $\Delta v = 100$~\kms, which is representative of the observed velocity dispersion in the emitting region (see Figure~\ref{fig:radex_spectra}). Molecular hydrogen was used as the collisional partner.

The modeled transitions included \COonezero, \COtwone, \COthreetwo, $^{13}$CO(3--2), and \COsixfive. The non-LTE grid was constrained using representative spectra in each line for the shell that were constructed by taking the mean value at each velocity over all pixels within the bubble aperture (Figure~\ref{fig:radex_spectra}). Brightness temperatures were determined for the $^{12}$CO transitions around 250--260~\kms (1--0: 42.1~K; 2--1: 27.7~K; 3--2: 36.6~K; 6--5: 16.3~K). The optical depth of the 3--2 line was constrained by a measured $^{12}$CO/$^{13}$CO(3--2) line ratio of 5.06. We adopt a $^{12}$C/$^{13}$C abundance of 25, consistent with measurements from CO by \citet{Martin+2019} and measurements using higher dipole species \citep{Martin+2021}. This corresponds to an effective optical depth of $\tau \approx 5.5$ for the \COthreetwo transition. To more conservatively isolate the molecular gas properties just at the velocities where the shell appears less contaminated by nearby emission (25--198~\kms), a second non-LTE analysis was performed using a a representative spectrum over the limited velocity range of Figure~\ref{fig:moment0_map} where the mean value at each velocity was measured for all pixels with emission in a moment 0 map constructed over this limited velocity range. Brightness temperatures were determined for the $^{12}$CO transitions around 185--195~\kms (1--0: 3.5~K; 2--1: 4.4~K; 3--2: 10.7~K; 6--5: 6.4~K). The optical depth constraint for this run was $\tau = 1.5$ for the \COthreetwo transition. Fitting was performed over the same parameter grid and using the same methodology as the first run.

\begin{figure}[htbp]
    \centering
    \includegraphics[width=\linewidth, trim={22cm 1cm 22cm 1cm}, clip]{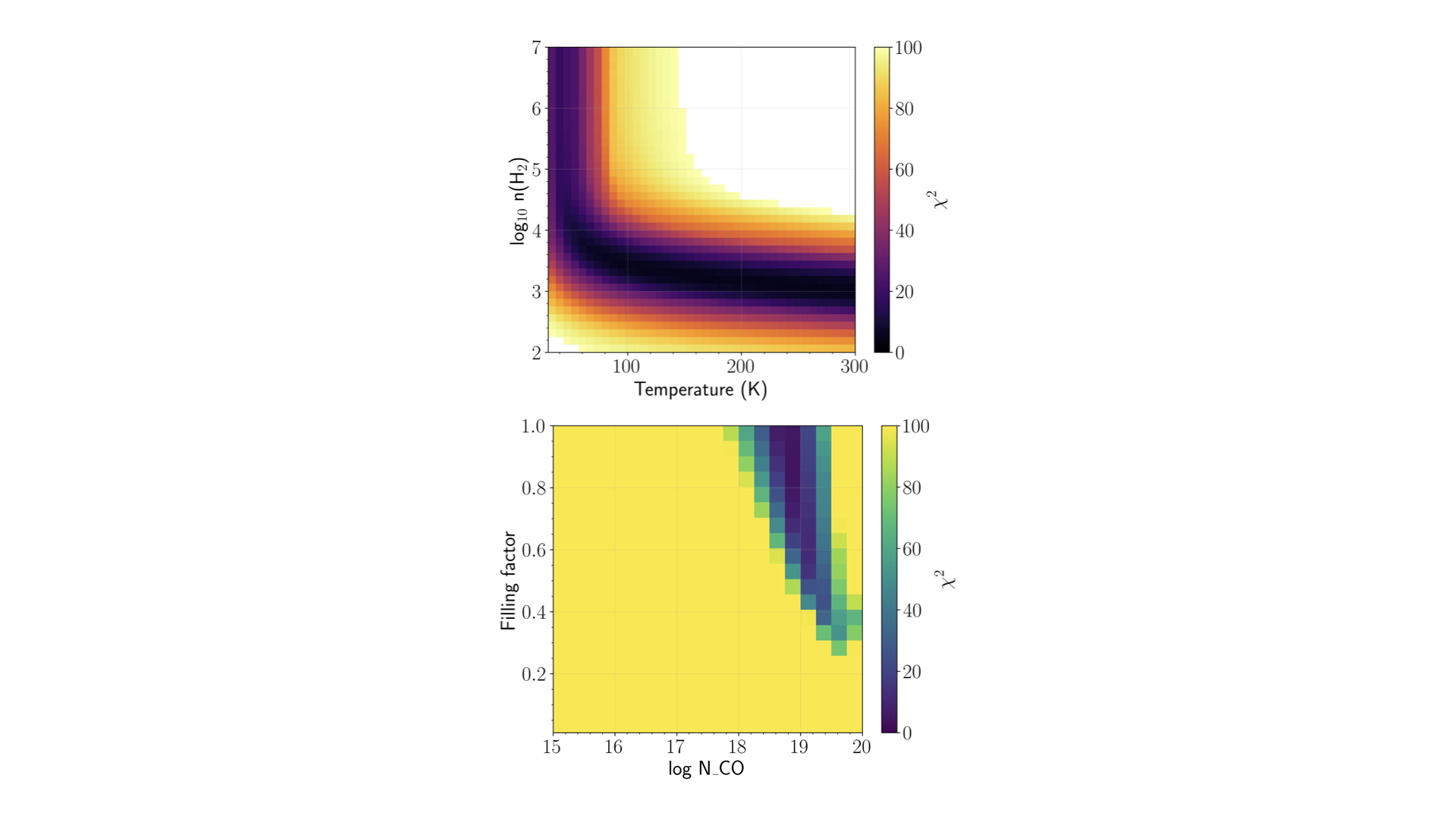}
    \caption{Reduced $\chi^2$ maps from the \texttt{RADEX} modeling using integrated intensities measured over the full CO emission profile. The top panel shows the minimum reduced $\chi^2$ as a function of kinetic temperature and \htwo density. The modeling constrains the gas density to $n({\rm H}_2) \gtrsim 10^3$~\cm, while the kinetic temperature remains largely unconstrained over the explore range. The bottom panel shows the minimum reduced $\chi^2$ as a function of CO column density and beam filling factor. A well-defined minimum is found at $N_{\rm CO} \approx 8.9 \times 10^{18}$~\cmtwo, with the elongated valley reflecting the degeneracy between CO column density and filling factor.}
    \label{fig:radex_solution_run1}
\end{figure}

\begin{figure}[htbp]
    \centering
    \includegraphics[width=\linewidth, trim={22cm 1cm 22cm 1cm}, clip]{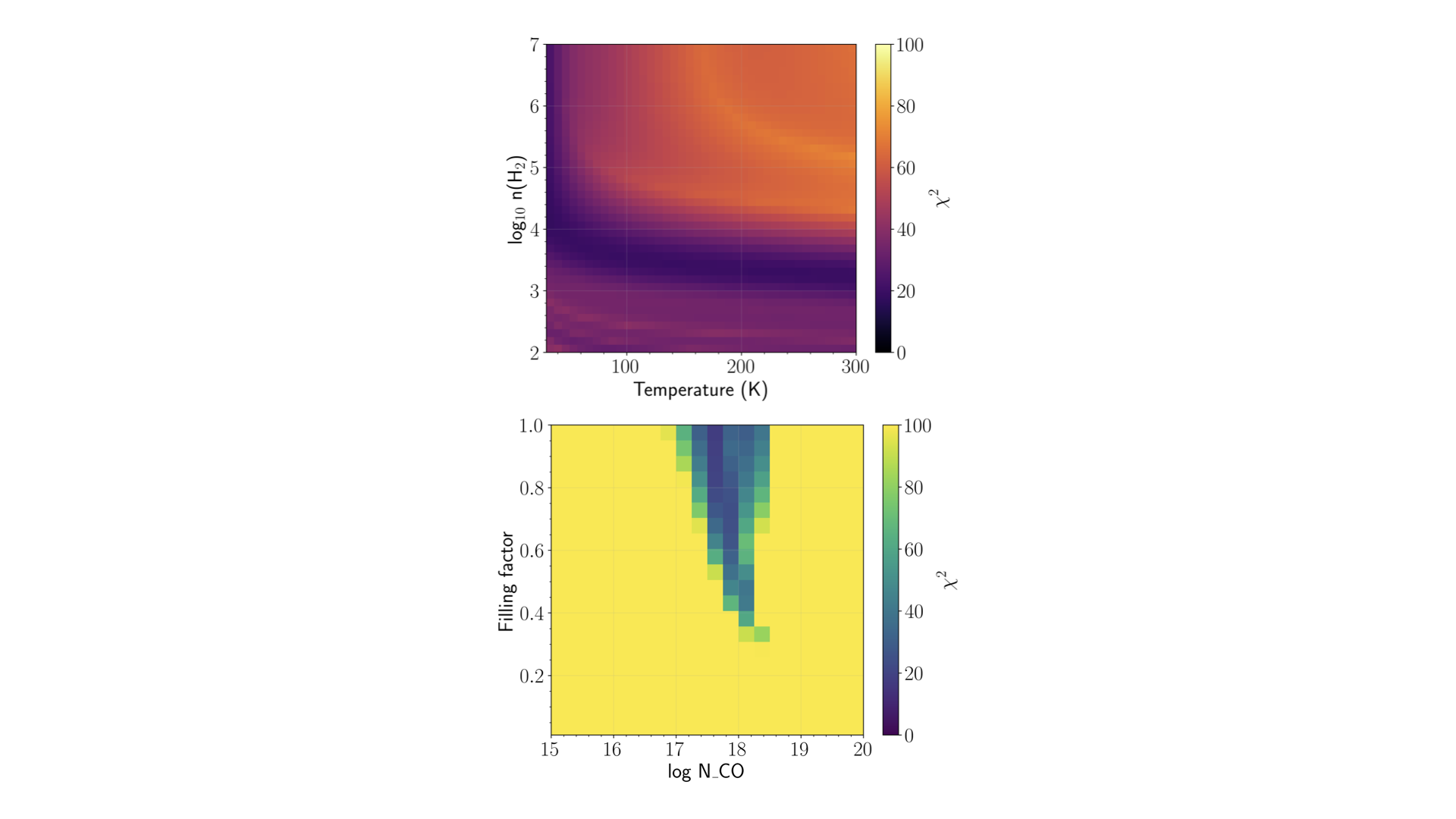}
    \caption{Reduced $\chi^2$ maps from the \texttt{RADEX} modeling using integrated intensities measured only over the velocity interval associated with the molecular shell. As in Figure~\ref{fig:radex_solution_run1}, the top panel shows the temperature--density parameter space and the bottom panel shows the CO column density--filling factor parameter space. The temperature is not well constrained, while acceptable solutions require densities of approximately $n({\rm H}_2) \gtrsim 10^3$~\cm. The best-fit CO column density is significantly lower than in the full-velocity model, with $N_{\rm CO} \approx 4.3 \times 10^{17}$~\cmtwo. The elongated minimum in the lower panel again illustrates the degeneracy between CO column density and beam filling factor.}
    \label{fig:radex_solution_run2}
\end{figure}

For all integrated intensities, we adopted the nominal ALMA flux calibration uncertainties for each observing band: 5\% for Band~3 (\COonezero), 10\% for Bands~6 and 7 (\COtwone, \COthreetwo, and $^{13}$CO(3--2)), and 20\% for Band~9 (\COsixfive), following the ALMA Cycle~13 Proposer's Guide \citep{ALMA_Cycle13_ProposersGuide}. These uncertainties were used to compute the reduced $\chi^2$ values. The reduced $\chi^2$ maps (Figures~\ref{fig:radex_solution_run1} and \ref{fig:radex_solution_run2}) demonstrate that acceptable solutions span a broad range of temperatures and densities. The kinetic temperature is essentially unconstrained, while densities of $n({\rm H}_2) \gtrsim 10^3$~\cm are preferred. In contrast the CO column density is relatively well constrained despite its degeneracy with the beam filling factor.

We find a peak CO column density of $N_{\rm CO} \sim 8.9 \times 10^{18}$~\cmtwo (bottom panel of Figure~\ref{fig:radex_solution_run1}). Assuming a CO abundance relative to molecular hydrogen of $X_{\rm CO} = 3 \times 10^{-4}$ \citep[e.g.,][]{Lacy+1994, Teng+2022, Teng+2023, He+2024}, we derive an \htwo column density of $N_{\rm H_2} \sim 2.9 \times 10^{22}$~\cmtwo. At a distance of 3.5~Mpc and an emitting area of $\sim 415$~pc$^2$ (which accounts for all pixels within the bubble aperture), our first model (incorporating emission at all velocities in the direction of the shell) yields a total molecular gas mass of $M_{\rm gas} \approx (2.8 \pm 0.8) \times 10^5$~\msun. We consider this mass an upper limit of the molecular gas content in the bubble as it likely includes foreground emission corresponding to gas orbits on a ring in the inner 200~pc of the galaxy \citep{Levy+2022}.

The same analysis applied for the limited velocity range yields $N_{\rm CO} \sim 4.3 \times 10^{17}$~\cmtwo (bottom panel of Figure~\ref{fig:radex_solution_run2}), corresponding to $N_{\rm H_2} \sim 1.4 \times 10^{21}$~\cmtwo. Across the same emitting area, the total molecular gas mass is $M_{\rm gas} \approx (1.3 \pm 0.3) \times 10^4$~\msun. We consider this mass a lower limit on the molecular gas content in the bubble because it misses the shell emission at higher velocities. Together, these measurements define the plausible molecular gas mass range that we attribute to the bubble/shell structure.

\subsection{Luminosity and Ionizing Flux\label{sec:lum_and_Q0}}

To quantify the energetic output associated with the bubble and its central source, we measure the line luminosities of the hydrogen recombination line \hfivea (Pfund~$\alpha$). This provides a direct constraint on the ionizing photon production rate ($Q_0$) of the embedded stellar population. This ionizing photon rate reflects the combined contribution of massive stars within SSC~6 and provides an independent constraint on the cluster's stellar content.

We calculate an \hfivea recombination line flux of $2.6 \times 10^{-13}$~erg~\persec~\cmtwo and line luminosity of $3.8 \times 10^{38}$~erg~\persec for the entire bubble region. We apply an extinction correction using an extinction map derived from the 9.7~$\micron$ silicate absorption feature using \texttt{PAHFIT} \citep{Smith+2007} assuming a mixed dust geometry (see U.~Siwakoti et al.\ in prep) with an average extinction of $A_{\rm H5\alpha} = 0.30$~mag within the bubble aperture (corresponding to a visual extinction of approximately 14.2~mag). Assuming a foreground screen geometry, the extinction-corrected luminosity is $L_{\rm H5\alpha} \sim 5.1 \times 10^{38}$~erg~$s^{-1}$.

The line luminosity is used to estimate the ionizing photon rate, $Q_0$, under the assumption of Case~B recombination. For an optically thin source, the ionizing flux is related to the recombination-line luminosity via

\begin{equation}
    Q_0 = \frac{L_{\rm H5\alpha}}{h \nu_{\rm H5\alpha}} \, \frac{\alpha_{\rm B}}{\alpha_{\rm H5\alpha}^{\rm eff}} ,
\end{equation}

\noindent where $\nu_{\rm H5\alpha} = 7.3 \times 10^{13}$~Hz is the transition frequency, $\alpha_{\rm B} = 2.6 \times 10^{-13}$~cm$^3$~\persec is the Case~B recombination coefficient, and $\alpha_{\rm H5\alpha}^{\rm eff} = 1.2 \times 10^{-14}$~cm$^3$~\persec is the effective recombination coefficient for the \hfivea transition \citep{Draine2011}. Recombination coefficients are adopted for an electron temperature $T_{\rm e} \sim 10^4$~K and density $n_{\rm e} \sim 10^4$~\cm \citep{Mills+2021}. The adopted coefficients vary by less than a factor of 2 over $T_e \sim 5000$--$2 \times 10^4$~K and are insensitive to density for this low-$n$ transition \citep{Storey+1995, Salgado+2017}.

For the entire bubble region, we find an ionizing photon rate of $Q_0 = 2.3 \times 10^{52}$~\persec. This is larger than the ALMA measurement of $Q_0 = 4 \times 10^{51}$~\persec, which was computed just for a 0\farcs18 aperture centered on SSC~6, as the ALMA recombination line data were not sensitive enough to recover the lower surface-brightness emission from the shell. Using a smaller-aperture radius for the more compact \hii region structure around SSC~6, equivalent to the resolution of the MIRI data (0\farcs35), yields $Q_0 = 8.7 \times 10^{51}$~\persec though this is likely an underestimate due to the MIRI beam response. This is more comparable to the ALMA-derived value, though we emphasize that both are underestimates of the full ionizing flux of SSC~6 if this source is indeed the primary ionizing source of the bubble studied here.  

\subsection{Theoretical Energy Requirements\label{sec:energy_calcs}}

To quantify the energy required to drive the observed bubble expansion, we compute the kinetic energy, momentum, and characteristic age using the expansion velocity measured from our forward modeling (Section~\ref{sec:forward_modeling}) together with the molecular gas masses derived from our non-LTE analysis (Section~\ref{sec:radex_modeling}).

The molecular gas provides the most robust tracer of the shell’s swept-up material because it is well constrained observationally and is spatially coincident with the bubble rim. While additional swept-up mass may exist in atomic and ionized phases, these components are not well constrained in our data and therefore not included. Our estimates should thus be interpreted as conservative lower limits on the total mass participating in the expansion. 

We adopt two molecular mass estimates from Section~\ref{sec:radex_modeling}. The lower value, $M_{\rm lower} = (1.3 \pm 0.3) \times 10^4$~\msun, is derived from the shell-associated CO velocity component alone, while the upper value, $M_{\rm upper} = (2.8 \pm 0.8) \times 10^5$~\msun, includes all molecular gas projected within the bubble aperture. We derive a characteristic age of $t_{\rm age} = 0.1 \pm 0.1$~Myr. The lower and upper mass estimates correspond to kinetic energies of $(1.1 \pm 1.0) \times 10^{51}$ and $(2.3 \pm 2.3) \times 10^{52}$~erg, respectively, and momenta of $(1.2 \pm 0.6) \times 10^6$ and $(2.5 \pm 1.4) \times 10^7$~\msun~\kms, respectively. These values are summarized in Table~\ref{tab:bubble_properties}.

\begin{table}[htbp]
\begin{threeparttable}
\caption{Properties of the bubble.}
\label{tab:bubble_properties}
\small
\setlength{\tabcolsep}{4pt}
\begin{tabular}{@{}l l@{}}
\toprule

\textbf{Property} & \textbf{Value} \\
\midrule
RA, $\alpha$ & 00\h 47\m 33.02\s \\
Dec, $\delta$ & $-25^\circ$ 17\arcmin 19\farcs57 \\
Radius & \\
\quad $r$ [arcsec] & $0.68 \pm 0.20$ \\
\quad $r$ [pc] & $11.5 \pm 3.4$ \\
$v_{\rm sys}$\textsuperscript{a} [\kms] & $231 \pm 28$ \\
$v_{\rm exp}$\textsuperscript{a} [\kms] & $90 \pm 44$ \\
$t_{\rm age} = r/v_{\rm exp}$ [Myr] & $0.1 \pm 0.1$ \\
$M_{\rm lower}$ [\msun] & $(1.3 \pm 0.3) \times 10^{4}$ \\
$M_{\rm upper}$ [\msun] & $(2.8 \pm 0.8) \times 10^{5}$ \\
$E_{\rm lower}$ [erg] & $(1.1 \pm 1.0) \times 10^{51}$ \\
$E_{\rm upper}$ [erg] & $(2.3 \pm 2.3) \times 10^{52}$ \\
$p_{\rm lower}$ [\msun\,\kms] & $(1.2 \pm 0.6) \times 10^{6}$ \\
$p_{\rm upper}$ [\msun\,\kms] & $(2.5 \pm 1.4) \times 10^{7}$ \\

\bottomrule
\end{tabular}
\begin{tablenotes}
\footnotesize 
\item \textsuperscript{a} The bubble's systemic and expansion velocity are refined via forward modeling of the PV structure (Section~\ref{sec:forward_modeling}).
\end{tablenotes}
\end{threeparttable}
\end{table}

\section{Discussion\label{sec:Discussion}}

The observations presented in this work reveal a compact bubble of ionized and molecular gas in the nuclear region of NGC~253, roughly centered on a massive cluster and compact \hii region (SSC~6) detected at wavelengths from near-infrared to radio continuum. The bubble exhibits a morphologically similar structure in warm and cool molecular gas and ionized gas. We now explore the nature of the source or sources driving the bubble expansion, the bubble’s evolution as an early-stage superbubble, and its connection to larger-scale nuclear outflows.

\subsection{Constraints on the Bubble's Driving Source\label{sec:driving_nature}}

\subsubsection{Properties of the Central Cluster\label{sec:cluster_properties}}

As noted in Section~\ref{sec:HST_data}, when comparing archival HST data of the NGC~253 nuclear region with our new JWST data we find a spatial offset compared to the previous adopted astrometry in \cite{Leroy+2018}. The identification of SSC~6 as the counterpart to the near-infrared cluster, and the non-detection of any continuum emission from SSC~5, is more consistent with the cluster properties derived by ALMA. \citet{Leroy+2018} find that SSC~5 is still heavily embedded, with a gas mass ($10^{5.3}$~\msun) comparable to its stellar mass ($10^{5.4}$~\msun), while SSC~6 has a much lower gas mass ($10^{3.6}$~\msun) relative to its stellar mass ($10^{5.3}$~\msun). SSC~6 also appears kinematically associated with the bubble. We find comparable systemic velocities for the bubble ($\sim 231 \pm 28$~\kms; Section~\ref{sec:kinematic_fit}) and SSC~6 \citep[$\sim 234$~\kms;][]{Mills+2021}, while SSC~5 has a lower systemic velocity \citep[though still within error; $\sim 210$~\kms;][]{Mills+2021, Levy+2021}.

While we identify SSC~6 as the near-infrared detected cluster, SSC~5 and SSC~7 are both also projected to lie inside the confines of the bubble. Continuum emission from SSC~5 is not detected in our JWST observations, however it is still possible that the cluster could have broken out of the back side of the obscuring clump and be contributing to the bubble's ionization. SSC~5 is also one of only three of the ALMA-detected SSCs which is observed to have a molecular outflow \citep{Levy+2021}. However, these outflows appear to be associated with clusters that have not yet cleared their immediate environment, making it less likely that SSC~5 is playing a role in evacuating the bubble cavity. In addition, we find that SSC~7 may not be a SSC at all, consistent with a suggestion by \citet{Levy+2022}. The ALMA SSCs were initially identified as compact sources with roughly co-spatial ionized gas and dust emission, however \citet{Levy+2022} do not detect compact dust continuum structure toward this cluster in higher-resolution ALMA observations. Its location at the edge of the JWST bubble and systemic velocity of $\sim 250$--270~\kms \citep{Mills+2021, Levy+2021}, makes it likely that this source is an over-density in the overall bubble structure as opposed to a compact \hii region ionized by a separate SSC.

The ionizing photon rate for SSC~6, derived from H4$\gamma$ (i.e., Brackett-$\gamma$, $\lambda=2.16$~\micron) by \citet{Kornei+2009}, is nearly twice the value derived by \citet{Bendo+2015} from H40$\alpha$ for the entire starburst. We therefore compare $Q_0$ estimates across different aperture sizes and diagnostics in Table~\ref{tab:Q0_methods} to attempt to determine an accurate $Q_0$ and mass for this cluster.

\begin{table*}[htbp]
    \centering
    \caption{Ionizing photon rates for NGC~253 in the literature.}
    \begin{tabular}{c c c c}
        $\log Q_0$\textsuperscript{a} (s$^{-1}$) & Method & Aperture Size & Source\\
    \hline
     \multicolumn{4}{l}{Entire Nucleus}\\
    \hline
        53.53 & H40$\alpha$ & $20'' \times 10''$ & \cite{Bendo+2015} \\
        53.48 & RRLs  & $17''$ beam & \cite{Puxley+1997}\\
        53.02 & Ne~II & $15''$ beam & \cite{Engelbracht+1998}\\
        52.91 & H4$\gamma$ & $15''$ beam & \cite{Engelbracht+1998} \\
        53.06 & H3$\alpha$ & $3.9'' \times 4.8''$ & \cite{AlonsoHerrero+2003} \\
        53.08 & 33 GHz  & 14 SSCs ($0.1-0.25''$ each) & \cite{Leroy+2018} \\
         \hline
        \multicolumn{4}{l}{Bubble}\\
        \hline
        52.36 & \hfivea & $0.68''$ & This work \\
         \hline
        \multicolumn{4}{l}{SSC 6}\\
        \hline
        53.77 & H4$\gamma$  & IR cluster & \cite{Kornei+2009}\\
        51.8-52.3 & H73$\alpha$, H92$\alpha$ & Radio Nucleus ($0.3''$) & \cite{Mohan+2002} \\
        51.94 & \hfivea & SSC~6 ($0.35''$) & This work \\
        51.60 & H40$\alpha$ & SSC~6 ($0.18''$) & \cite{Mills+2021} \\
    \end{tabular}
    \label{tab:Q0_methods}
    \vspace{2mm}
    
    \footnotesize
    \textsuperscript{a} All measurements have been distance-corrected to 3.5~Mpc and are ordered approximately from largest to smallest aperture size.
\end{table*}

We first compare to values from \citet{Mills+2021} who also use H40$\alpha$ to estimate $Q_0 \sim 4.0 \times 10^{51}$~\persec) in a compact 0\farcs18 aperture centered on the cluster. This estimate assumes no ionizing photons escape; since SSC~6 likely ionizes the bubble structure, it should be considered a lower limit on the cluster properties. We measure $Q_0 = 8.7 \times 10^{51}$~\persec (extinction-corrected assuming a foreground screen geometry; $A_{\rm H5\alpha} \sim 0.24$~mag) from \hfivea emission in a similarly compact aperture (0\farcs35) centered on SSC~6, in good agreement with this ALMA-based estimate. For the entire bubble ($r_0 =$ 11.5~pc = 0\farcs678), we estimate $Q_0 = 2.3 \times 10^{52}$~\persec ($A_{\rm H5\alpha} \sim 0.30$~mag; $A_V \sim 14.2$~mag). This is more than an order of magnitude lower than the \citet{Kornei+2009} estimate despite being measured over a comparable aperture. We are unable to determine the cause of this discrepancy. Future JWST observations of the H4$\gamma$ line should be able to reveal if this was an error in calibration or the assumed extinction ($A_V = 17.7$~mag). 

To calculate the cluster mass then requires assuming a cluster age and initial mass function (IMF). The timescale for massive stellar populations to remain embedded in their natal cloud has a typical duration of $\sim 3$~Myr \citep{Kim+2021}, whereafter they emerge and become visible at optical and infrared wavelengths. The absence of He-recombination line emission in SSC~6 \citep{Mills+2021} also suggests an age greater than $\sim 3$~Myr \citep[whereafter He-ionizing radiation decreases steeply;][]{Levesque+2013}. We therefore place a 3~Myr lower limit on SSC~6's age. Our current observations do not yield strong constraints on the upper age limit of the cluster, so we include a larger age estimate of 5.7~Myr from \citet{Kornei+2009} for calculating the possible mass range for this source. Using the \texttt{pySTARBURST99} framework \citep{Hawcroft+2025}, our estimated ionizing flux for the entire bubble, and assuming a Kroupa IMF with a maximum mass of 100~\msun and a minimum mass of 0.1~\msun, we then estimate the mass of SSC~6. For a zero-age main sequence stellar population, as previously assumed by \citet{Leroy+2018} and \citet{Mills+2021} we would find a mass of $7 \times 10^5$~\msun. Adopting our 3~Myr lower age limit, we find $1.2 \times 10^6$~\msun. At 5.7 Myr (the \citealt{Kornei+2009} adopted age), our estimate increases to $3 \times 10^7$~\msun. While we believe that this larger mass is unlikely, we currently do not have good constraints on the upper age limit for SSC~6.  

We also compare the cluster mass ($M_{\rm cl} \sim 1.2 \times 10^6$~\msun, assuming an age of 3~Myr) to estimates of the molecular masses associated with the bubble and shell of $(1.3 \pm 0.3) \times 10^4$ to $(2.8 \pm 0.8) \times 10^5$~\msun, calculating $M_{\rm cl} / (M_{\rm bubble}+M_{\rm cl}) \sim 0.81$--$0.99$. If the observed bubble is the remnants of the natal cloud of SSC~6, this ratio would imply an unusually high local star formation efficiency. One explanation for this is that the bubble mass is dominated by ionized and/or atomic gas. Alternatively or additionally, SSC~6 could have already dispersed or decoupled from its natal cloud, as is seen for example in the less-massive \citep[$\sim 2 \times 10^4$~\msun;][]{Espinoza+2009} 2--3 Myr-old Arches cluster in the Milky Way Center \citep{Figer+2002, Clark+2018}, which appears to exist in a largely evacuated giant \hii region with radius $\sim 15$~pc \citep{Lang+2001, Hankins+2017}. Recent studies suggest that more massive clusters may emerge from their natal clouds even more quickly than less massive clusters \citep{Pedrini+2026}. This would be consistent with observations of compact molecular outflows in several of the embedded clusters in NGC~253, which have mass loss rates sufficient to disrupt their natal cloud in as little as 0.01--0.3~Myr \citet{Levy+2021}. 
If SSC~6 is already separated from its host cloud, the bubble mass might then be dominated by swept-up ambient material. Taking the bubble radius and assuming a uniform ambient density of $2 \times 10^3$~\cm would yield a mass of $\sim 4.4 \times 10^5$~\msun, approximately $1.5\times$ the value of our largest estimated mass. This suggests that the shell may indeed be formed from swept-up ambient material.

\subsubsection{An X-Ray Counterpart \label{sec:cluster_xray}}

\begin{figure}[htbp]
    \centering
    \includegraphics[width=0.47\textwidth, trim={15cm 0cm 15cm 0cm}, clip]{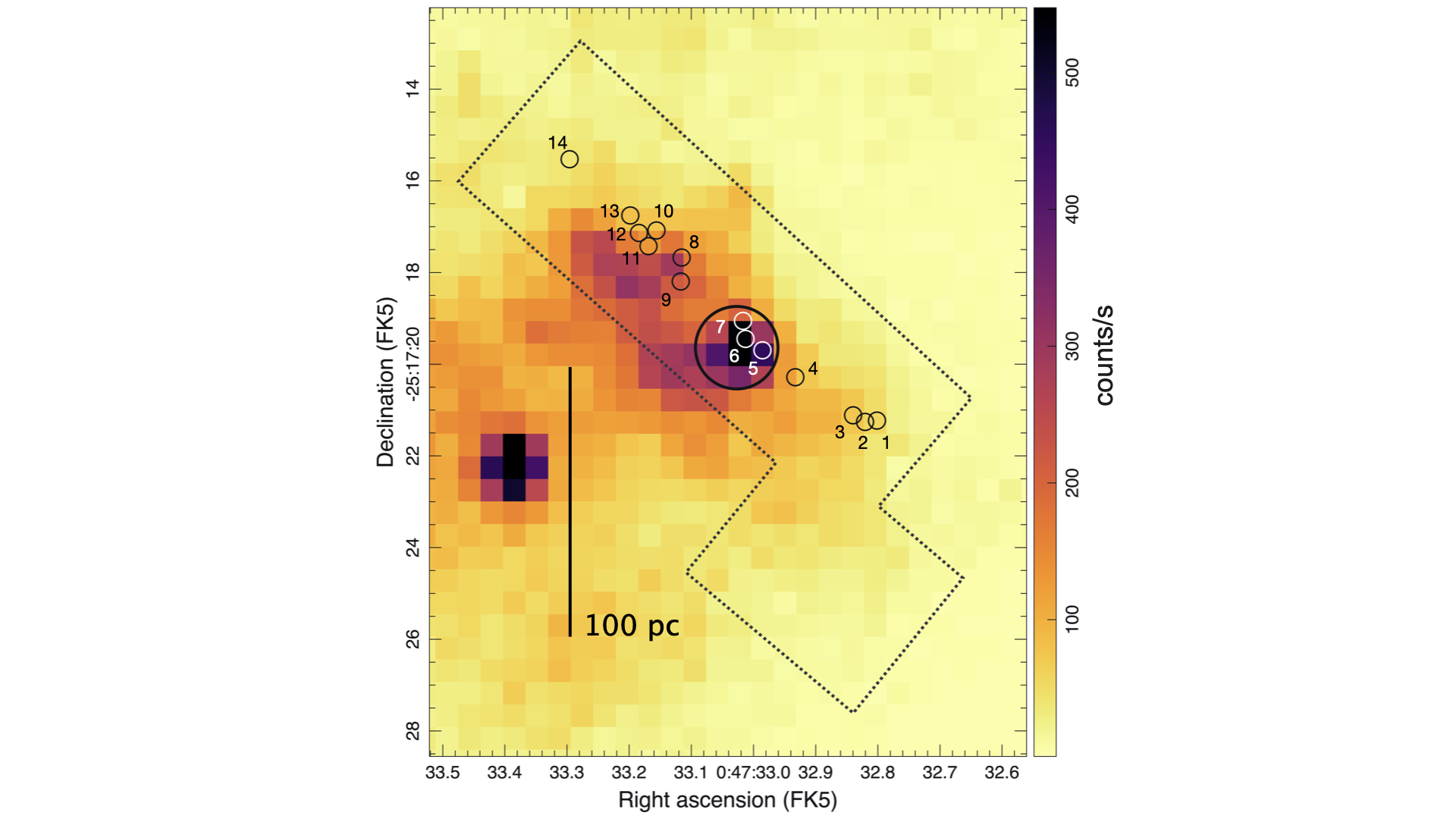}
    
    \caption{Comparison of the bubble position (black circle) overlaid on Chandra 0.5 -- 7~keV data from \citet{Lopez+2023}. Positions of all the ALMA SSCs from \citet{Leroy+2018} are shown as small circles. The MIRI-MRS Channel 1 footprint is overlaid as a black dotted region.}
    \label{fig:xray}
\end{figure}

High-resolution Chandra observations reveal a luminous X-ray point source spatially coincident with the infrared bubble and SSC~6 (Figure~\ref{fig:xray}). This point source corresponds to Source~A in \citet{Lehmer+2013} and X-2 in \citet{Muller-Sanchez+2010}, where it is identified as the brightest soft X-ray source in the central 100~pc of NGC~253. \citet{Muller-Sanchez+2010} report a hard X-ray luminosity of $L_{2-10~\mathrm{kev}} = 1.2 \times 10^{38}$~erg~\persec (distance-corrected). 

\cite{Gunthardt+2015} also note the coincidence of the infrared and X-ray emission, and argue that the infrared cluster (SSC~6) is the most likely candidate for the nucleus of NGC~253, given that it is the brightest source in the near and mid-IR (though they note that there are no signatures of an active galactic nucleus (AGN) in the infrared line emission). They use observations of the CO $v = 2$--0 bandhead to infer the coexistence of stellar populations older than a few $\times 10^7$~yr together with a younger population of a few Myr, which they attribute to prolonged or episodic star formation at the core of the galaxy. However, this CO feature is also seen up to 100~pc away from the position of SSC~6 and is ubiquitous in starbursts \citep{Armus+1995, Oliva+1995, Buiten+2024}. In contrast to this interpretation, \citet{Levy+2022} find that SSC~6 has kinematics consistent with lying on a ring of other young embedded clusters, orbiting at a radius of $\sim 100$~pc. This ring is centered on the dynamical center of \citet{Anantharamaiah+1996} which is determined from the kinematics of the ionized gas traced by radio recombination lines. Our JWST observations support this picture, in which SSC~6 is one of many SSCs in the nucleus, but appears as the brightest source at near-infrared and MIR wavelengths because it corresponds with or has created a hole in the overall extinction toward this region.

The hard X-ray luminosity from \citet{Muller-Sanchez+2010} is roughly two orders of magnitude higher than typical supernova remnants (SNR), which have X-ray luminosities of $10^{34}$--$10^{36}$~erg~\persec \citep{Oskinova2005}. Source~A has been detected over more than two decades with relatively stable luminosity \citep[e.g.,][]{Strickland+2000, Muller-Sanchez+2010, Lehmer+2013}, which further rules out individual SNR or transient explosive events, both of which would fade on timescales of years to centuries. Unlike an AGN, broadband NuSTAR and Chandra observations show that Source~A lacks a characteristic flat power-law spectrum \citep{Lehmer+2013, Lehmer+2019}. Additionally, consistent with \citet{Gunthardt+2015}, no high-ionization line emission has been found at this position in the JWST data (U.~Siwakoti et al.\ in prep) that would be consistent with an AGN.

We are left with several possibilities. First, the X-ray luminosity could be due to Wolf-Rayet (WR) stars. The brightest individual WR systems (colliding-wind binaries) have X-ray luminosities of $10^{34}$--$10^{35}$~erg~\persec \citep{Pollock+2018, Arora+2020}. The less massive Galactic Center Arches cluster ($\sim 2 \times 10^4$~\msun) hosts 13 WR stars \citep{Martins+2008} and has an X-ray luminosity of $4 \times 10^{35}$~erg~\persec \citep{Yusef-Zadeh+2002}. Based on these estimates SSC~6 would need to host 1000--10000 WR stars to account for the luminosity of the coincident X-ray source. As our \texttt{pySTARBURST99} modeling indicates that for a cluster mass of $1.2 \times 10^6$~\msun we would expect to have 5400 stars with masses above 15~\msun (which could be either O or WR stars), this would require a ratio of WR/(WR+O) stars to be between 0.2--1. Such high ratios can be reached in young starburst systems; modeling of \citet{Schaerer+1998} finds that the ratio reaches 1 at 4.5 Myr for solar-metallicity populations and at 3.5 Myr for supersolar (Z=0.04) metallicities. This would also be evidence for a $>$3~Myr age for SSC~6, but such a simple scaling may overestimate the X-ray flux. Modeling by \citet{Oskinova2005} estimates that the peak X-ray luminosity of a $10^6$~\msun cluster (occurring at an age of $\sim 3.5$~Myr and resulting from WR winds) is only $\sim 10^{37}$~erg~\persec, an order of magnitude less than the observed value. For WR stars to account for the X-ray luminosity thus likely requires a cluster mass significantly in excess of $10^6$~\msun.  

Second, we could seek to explain the luminosity with supernovae. While a single SNR is not luminous enough to explain the X-ray emission, a naive scaling of a larger number of SNRs (100--10000, based on individual source luminosities) could explain the total flux. For our estimated cluster ages of 3--5.7 Myr (and masses of $1.2 \times 10^6$--$3 \times 10^7$~\msun), the SN rate derived by our \texttt{pySTARBURST99} modeling ranges from $5 \times 10^{-4}$ to $1.4 \times 10^{-2}$ per year. This implies $\sim 50$--1400 SN could have taken place over the 0.1~Myr lifetime of the bubble. Again, this is likely an optimistic estimate of their contribution to the X-ray flux; modeling of the X-ray emission from a $10^6$~\msun cluster by \citet{Oskinova2005} found that the SNe-driven wind is comparably luminous to the WR driven wind ($\sim10^{37}$~erg~\persec) and decreases over time. For SNe to account for the X-ray luminosity then also likely requires a cluster mass significantly in excess of $10^6$~\msun. 

Alternatively, the spectral properties of the Chandra source, including the hard X-ray detection, steep spectral slope, and persistent luminosity, could be consistent with a single luminous X-ray binary source, likely a high-mass X-ray binary (HMXB), embedded within the nuclear starburst. HMXBs also require the prior formation of a compact object through a SN. Assuming that the first SNe occur when the cluster is around 3~Myr \citep{Badmaev+2024}, both of these scenarios are consistent with our independent age estimates for SSC~6 of at least 3--5~Myr. We have also searched for additional evidence of SNe in the bubble by examining its radio spectral index. \citet{Mills+2021} find $-0.5 \lesssim \alpha \lesssim 0$, inconsistent with purely non-thermal synchrotron emission and instead suggesting a mixture of thermal free-free and non-thermal components. While this is unlikely to be the case if the bubble is powered by hundreds to thousands of recent SNe, it could be consistent either with a few SNe or the presence of strong WR winds. We note that, given the moderate ($\sim$0\farcs5) Chandra spatial resolution and the possibility of multiple X-ray binaries in the nuclear region, a chance alignment between Source~A and SSC~6/the infrared bubble cannot be ruled out. However, across the entire nuclear region only three X-ray sources are comparably bright, and at least one of these is known to be strongly variable, unlike Source~A. Given the small number of such luminous sources, the probability of a chance positional coincidence with SSC~6 is correspondingly low.

\subsubsection{Driving Mechanism \label{sec:driving_mechanism}}

Our estimates of the bubble kinetic energy range from $(1.1 \pm 1.0) \times 10^{51}$ to $(2.3 \pm 2.3) \times 10^{52}$~erg. We evaluate three possible energy sources: stellar winds under classical, energy-conserving assumptions; stellar winds under a momentum-conserving, radiatively inefficient framework motivated by recent theoretical work; and supernovae under a cooling-modified framework.

With the classical \citet{Weaver+1977} framework, we calculate the continuous energy injection required from stellar winds (total mechanical energy) to create the bubble as $E_{\rm tot} \sim 1.8 \times 10^{53}$~erg (assuming an ambient molecular gas density of $n_0 = 2 \times 10^3$~\cm), using the closed-form expression $E_{\rm tot} = (10 \pi / 3) \rho_0 v_{\rm exp}^2 R^3$ from \citet{Sakamoto+2006}, in which $\sim 20\%$ of the injected mechanical energy resides in the shell's kinetic energy at a given time under the \citet{Weaver+1977} similarity solution. This is the energy required under the classical energy-conserving assumption, and is naturally larger than the bubble kinetic energies in Table~\ref{tab:bubble_properties} (which reflect only the shell's present kinetic energy, not total energy injected). WR stars are expected to be present in young clusters between 3--4 and 5--6~Myr \citep{Leitherer+1999,Hadfield_Crowther2007}, with stellar wind pressure a likely driver of the bubble for our assumed age range for SSC~6. WR stars are predicted to produce mechanical luminosities up to 1--$2 \times 10^{40}$~erg~\persec for a $10^6$~\msun cluster \citep{Leitherer+1999}. Integrated over the bubble lifetime ($\sim 0.1$~Myr), this yields wind energies of $\sim (3$--$6) \times 10^{52}$~erg. This is within a factor of 3--6 of the inferred injection energy of $1.8 \times 10^{53}$~erg, indicating that stellar winds from WR stars could could make a substantial contribution to driving the bubble.

If radiative cooling is significant (e.g., due to mixing at the bubble interior) the bubble may instead be in a momentum-conserving phase, in which the relation between injected wind energy and bubble kinetic energy differs from the \citet{Weaver+1977} scaling, and a lower mechanical energy input could sustain the observed expansion \citep{Lancaster+2021a, Lancaster+2021b}. \citet{Lancaster+2021a} proposed a new theoretical model for the expansion of stellar wind bubbles under conditions present in turbulent molecular clouds (with very high surface densities), appealing to efficient energy loss through turbulent, fractal, radiative mixing layers between the hot shocked wind and surrounding cloud. In this ``efficiently cooled" limit, the shell momentum is set directly by the wind momentum input rate, which scales linearly with the driving cluster's stellar mass. Applying this model to our lower and upper momentum estimates (Table~\ref{tab:bubble_properties}) yields corresponding cluster masses of $\sim 2 \times 10^6$ and $4 \times 10^7$~\msun, respectively, and an inferred injected wind energy of $\sim 6.3 \times 10^{52}$~erg. These cluster masses overlap with our independent \texttt{pySTARBURST99}-based cluster mass estimated for SSC~6 ($1.2 \times 10^6$~\msun at 3~Myr; $3 \times 10^7$~\msun at 5.7~Myr; Section~\ref{sec:cluster_properties}), and the inferred injected wind energy is consistent with the $\sim (3$--$6) \times 10^{52}$~erg expected from WR stellar winds over the bubble lifetime. This agreement supports stellar winds from SSC~6 as a viable power source for the bubble.

Clustered core-collapse SNe may contribute to driving the bubble, but are unlikely to dominate its present energetics. Applying the cooling-modified superbubble scaling of \citet{El-Badry+2019} suggests that $\sim 390$--27600 SNe would be required to supply our inferred kinetic energy, although this should be regarded as an order-of-magnitude estimate given that the model is extrapolated to the higher densities in SSC~6's environment. Even the lower end is comparable to or exceeds the $\sim 100$--3000 SNe permitted by our \texttt{pySTARBURST99} modeling over the bubble's lifetime. Thus, while we cannot rule out some contribution from SNe, stellar winds provide a plausible explanation for the bubble. Even if the present $\sim 0.1$~Myr-old bubble predates the onset of SNe, SSC~6 should enter a phase of additional mechanical energy injection from core-collapse SNe within the next few Myr.

The agreement between WR mechanical luminosities (integrated over the bubble lifetime) and injection energy with the \citet{Weaver+1977} and \citet{Lancaster+2021a} frameworks makes stellar winds the most likely driving mechanism for the observed bubble, though we caution that a modest supplementary SN contribution cannot be excluded. Conclusively answering this will require better estimates of the cluster age. Models \citep[e.g.,][]{Leitherer+1999} predict that WR stars dominate the wind at ages of 3--4~Myr before the onset of the first SNe, while the mechanical luminosity from SNe and WR stars are comparable between 4--6~Myr, and SNe are the dominant wind source after that. As powering the observed bubble by stellar winds requires a large WR to O star ratio, the detection of clear WR spectral features from the cluster and measurements of this ratio would also help determine its driving source.

\subsubsection{\texttt{TRINITY} Validation of Bubble Dynamics \label{sec:TRINITY}}

As an independent dynamical check on this picture, we ran the observed bubble parameters through \texttt{TRINITY} \citep{TRINITY+2026}, a Weaver-based bubble-evolution code that additionally accounts for gravity, radiation pressure, photoionized gas, and radiative cooling, using \texttt{pySTARBURST99} feedback tables (and therefore no SN contribution at $\sim 3$~Myr). We fix the shell radius at our observed value ($r_0 = 11.5$~pc), adopt an ambient density of $n_0 = 2 \times 10^3$~\cm, and use a cluster age of 3~Myr, consistent with our independently derived lower age limit for SSC~6 (Section~\ref{sec:cluster_properties}).

At our fiducial cluster mass ($1.2 \times 10^6$~\msun, from our $Q_0$-based estimate), \texttt{TRINITY} reproduces the observed radius, swept-up shell mass ((1.5--9.6)$\times 10^5$~\msun), kinetic energy ($1.2 \times 10^{52}$~erg), and momentum ($2.4 \times 10^7$~\msun~\kms) to within the measurement uncertainties (Table~\ref{tab:bubble_properties}). The shell retained $\sim$ 18\% of the injected mechanical energy (6.8 $\times 10^{52}$~erg), in good agreement with the $\sim$ 20\% expected under the classical \citet{Weaver+1977} energy-conserving solution. This injected wind energy is lower than our Weaver-based estimate of $1.8 \times 10^{53}$~erg (Section~\ref{sec:driving_mechanism}), but the two estimates assume different expansion velocities, so the discrepancy is expected. At our fiducial cluster mass and age, \texttt{TRINITY} predicts an expansion velocity of $v_{\rm exp} = 53$~\kms, which is less than (though still consistent within the uncertainty of) our best-fit value of $90 \pm 44$~\kms, and a kinematic age of $\sim$0.1~Myr, consistent with our observationally-derived dynamical age of the bubble. Matching our best-fit velocity exactly instead requires a higher cluster mass of $5.7 \times 10^6$~\msun, a factor of $\sim 4.8$ above the fiducial estimate but still below the cluster mass derived for the \citet{Kornei+2009} 5.7~Myr upper age limit ($3 \times 10^7$~\msun).

To place the bubble in a broader evolutionary context, we also ran \texttt{TRINITY} continuously forward in time at our fiducial cluster mass, illustrating the bubble's longer-term growth trajectory. These continuous runs reach $\sim$ 70~pc within 3~Myr of continued expansion. The high pressures of the NGC~253 nuclear disk \citep[$P/k \sim 10^{7-8}$K\cm;][]{Sakamoto+2006, Leroy+2018} may limit this growth and/or favor preferential expansion along the vertical direction.

Notably, the model indicates that radiation pressure contributes a non-negligible fraction of the feedback, accounting for $\sim$ 14\% at our fiducial cluster mass. The remaining $\sim$ 86\% is supplied by winds and photoionized gas pressure, suggesting radiation pressure may play a secondary but non-trivial role in the bubble's early dynamics. The models further indicate that the shell remains optically thick to ionizing photons throughout, consistent with our assumption of negligible LyC escape from the bubble (Section~\ref{sec:cluster_properties}). Taken together, these independent dynamical models support a picture in which pre-SN feedback (primarily stellar winds, supplemented by photoionization and radiation pressure) is sufficient to set the bubble's early evolution.

\subsection{The Evolution of a Nuclear Superbubble\label{sec:bubble formation}}

Recent simulations of the nuclear regions of Milky Way-like galaxies from \citet{Tress+2025} find that the short orbital times and strong shear in galaxy centers can rapidly decouple young stellar associations from their natal clouds, preventing radiation and SNe feedback from acting coherently. Feedback becomes distributed across the CMZ rather than concentrated, preventing the formation of large classical superbubbles and instead sustaining diffuse turbulence inefficient at disrupting dense gas.

The nuclear region of NGC~253 contains $\sim 10$ times the high-density gas mass and $\sim 3$ times the dense-gas mass fraction of the Milky Way center, with $3.5 \times 10^8$~\msun of \htwo, fueling a SFR of $\sim 2.8$~\msun~\peryr \citep{Ott+2005, Bendo+2015, Tanaka+2024} compared to Milky Way center values of  2--6$\times 10^7$~\msun and SFR of 0.09~\msun~\peryr \citep{Barnes+2017}. The Milky Way CMZ also contains several compact young massive clusters \citep[the `Arches' and `Quintuplet' clusters with ages 2.5~Myr and 4~Myr and masses $\sim 10^4$~\msun;][]{Figer+1999, Figer+2002, Espinoza+2009, Clark+2018}, though these are slightly older and less massive than the embedded population in NGC~253. Neither cluster is embedded, and they are believed to have decoupled from or dispersed their natal clouds. 

The bubble structure analyzed here provides a counterexample to the strictest interpretation of the \citet{Tress+2025} scenario, demonstrating that cluster feedback can organize and accelerate the surrounding ISM locally, even in a strongly sheared nuclear environment. We have suggested that relatively low bubble mass indicates that this structure may not represent the remnants of the SSC~6 natal cloud. Instead, it is likely ambient material that has been swept up by the passage of the cluster forming a shell of molecular material that surrounds the ionized bubble. This likely occurs due to a higher volume filling fraction of gas in the more massive central gas reservoir of NGC~253, compared to the simulations of the Milky Way center in \citet{Tress+2025}.

The bubble's evolutionary state can be characterized by comparing its expansion and cooling timescales. The expansion timescale, $t_{\rm exp} \sim R / v_{\rm exp} \sim 0.1 \pm 0.1$~Myr, provides a characteristic upper limit estimate of the bubble's dynamical age for a decelerating shell, while the radiative cooling time for the bubble's interior conditions ($T \sim 2 \times 10^7$~K and $n \sim 30$~\cm; from \texttt{TRINITY} modeling in Section~\ref{sec:TRINITY}) is $t_{\rm cool} \sim 1$~Myr \citep{Weaver+1977, Strickland+1999}. The hot interior can therefore remain approximately adiabatic over the present lifetime of the bubble, although the swept-up shell itself may already be radiative. As the bubble evolves, continued cooling and mixing at the interface between the hot interior and shell may increasingly reduce its expansion rate and promote fragmentation. Whether SSC~6 remains dynamically coupled to the expanding structure will determine its subsequent evolution. We measure a systemic velocity of $231 \pm 28$~\kms (Section~\ref{sec:kinematic_fit}), consistent with the $\sim 234$~\kms velocity measured for the compact \hii region associated with SSC~6 by \citet{Mills+2021}, providing no strong evidence for significant decoupling at present. SSC~6 is nevertheless offset by $\sim 3$~pc from the adopted bubble center, which could reflect a modest projected displacement or the earliest stages of dynamical separation driven by shear. If the cluster remains coupled, continued feedback may sustain the bubble for several Myr; if it decouples, cooling, compression, and shear could instead produce a fragmented or ``orphaned'' structure similar to those observed elsewhere in the NGC~253 nuclear disk \citep{Sakamoto+2006, Krieger+2019, Konishi+2025}.

\subsection{Coupling of Multiphase Gas to the Superwind\label{sec:coupling}}

Regarding blowout, the bubble shows evidence for partial breakout. The backwards ``C"-shaped or incomplete morphology suggests that the shell has ruptured along one side. Channel maps (Figure~\ref{fig:ArII_channel_maps}) also show extended emission towards the south-east, opposite the backwards ``C". The bubble's location within the dense nuclear disk and its modest size suggest it has not fully broken out of the dense molecular layer, but the ``C"-shaped morphology and minor-axis outflow indicate early-stage blowout, where feedback is beginning to create pathways for gas to escape the disk and feed the galactic-scale wind.

Measurements by \citet{Strickland+1999} for a cluster in NGC~5253 suggest that individual massive clusters in starburst galaxies are expected to drive localized feedback into the surrounding starburst region, even if enclosed by a larger superbubble. However, sufficiently massive clusters may confine their winds rather than drive strong localized outflows \citep{Turner+2003}. Depending on the spatial separation of clusters and the local star formation history, each cluster may temporarily inflate its own superbubble before neighboring bubbles interact and eventually merge. Our bubble measurements ($\sim 11.5$~pc radius, $\sim 0.1$~Myr age, $v_{\rm exp} \sim 90$~\kms) are consistent with a young, localized superbubble driven by an individual cluster before it merges with or contributes to the larger-scale outflow.

Finally, we compare this localized feedback event to the large-scale nuclear outflow. The molecular outflow contains $\sim 2$--$4 \times 10^7$~\msun \citep{Zschaechner+2018} and carries a total energy of order $10^{55}$~erg \citep{Bolatto+2013, Walter+2017}. The present bubble and shell therefore represent only a small fraction of the total wind energy budget, indicating that such feedback events are required over the lifetime of the starburst to power the galaxy-scale wind. The bubble thus likely represents one example of the localized feedback events that collectively power the galaxy-scale superwind.

\section{Conclusion\label{sec:Conclusion}}

In this paper, we present a multiphase analysis of a compact, expanding bubble in the nuclear starburst of NGC~253 using JWST MIRI-MRS spectroscopy and ALMA submillimeter observations. Our results provide new constraints on the origin, energetics, and environmental impact of localized stellar feedback at the scale of individual SSCs, and its potential connection to the galaxy-scale superwind. Our main conclusions are as follows:

\begin{enumerate}
	\item We identify a bubble-like structure with a radius of $\sim 11.5 \pm 3.4$~pc ($\sim 0\farcs68$) centered near SSC~6 \citep{Leroy+2018}. Molecular gas (\htwo and CO) is concentrated along the edges, and ionized emission (\arii and \hfivea) is stronger towards the interior. The structure is spatially and kinematically coherent across tracers, indicating a shared expansion pattern.
     
    \item We model a bubble expansion velocity of $90 \pm 44$~\kms. The young characteristic age of $\sim 0.1 \pm 0.1$~Myr and kinetic energy of $\sim 10^{51-52}$~erg are consistent with sustained feedback from stellar winds or potentially SNe from SSC~6. Independent \texttt{TRINITY} modeling shows that pre-SN winds and radiation pressure can reproduce the bubble's dynamics.
    
    \item The molecular mass of the bubble inferred from \texttt{RADEX} ranges $(1.3 \pm 0.3) \times 10^4$ to $(2.8 \pm 0.8) \times 10^5$~\msun. Given the small amount of molecular gas associated with SSC~6 and the fact that SSC~6 alone of the ALMA-detected clusters is visible in the near-infrared, SSC~6 is likely the most evolved of the clusters seen in the NGC~253 nucleus. The presence of X-ray emission, either from past supernovae or a large population of WR stars (Section~\ref{sec:driving_nature}) suggests that SSC~6 has an age $\gtrsim 3$~Myr.
    
    \item We measure an ionizing photon rate of $Q_0 = 2.3 \times 10^{52}$~\persec for the bubble region. Using \texttt{pySTARBURST99} models, we then estimate a cluster mass of $M_\star \sim 1.2 \times 10^6$~\msun for SSC~6 at an age of 3~Myr, containing approximately 5400 O-type stars and corresponding to an estimated core-collapse supernova rate of $\sim 6 \times 10^{-4}$~\peryr for a cluster age between 3--6~Myr. 

    \item The cooling timescale ($\sim 1$~Myr) exceeds the expansion timescale ($\sim 0.1$~Myr), suggesting that the hot bubble interior can remain approximately adiabatic over its current lifetime, although the swept-up shell may already be radiative. The backwards ``C"-shaped morphology and kinematic evidence for high-velocity material escaping along the minor axis of the bubble may indicate early-stage blowout in which sustained mechanical feedback is beginning to couple material to the larger-scale superwind.
\end{enumerate}

\section*{Acknowledgments}

The authors thank the anonymous referee for their constructive and helpful comments, which improved the manuscript.
The authors also thank Bret Lehmer and Bruce Draine for helpful conversations related to this work. 

E.A.C.\ Mills gratefully acknowledges funding from the National Science Foundation under Award Nos.\ 2206509 and CAREER 2339670. 
E.A.C.\ Mills, U.\ Siwakoti, A.D.\ Bolatto, J.D.\ Smith, Y.-H.\ Teng, S.A.\ Cronin and S.E.\ Duval acknowledge support from the Space Telescope Science Institute via grant No.\ JWST-GO-01701. 
V.V.\ acknowledges support from the Comité ESO Mixto 2024 and from the ANID BASAL project FB210003. 
R.H.-C.\ thanks the Max Planck Society for support under the Partner Group project ``The Baryon Cycle in Galaxies" between the Max Planck Institute for Extraterrestrial Physics and the Universidad de Concepción. R.H-C.\ also gratefully acknowledges financial support from ANID - MILENIO - NCN2024\_112 and ANID BASAL FB210003.
R.S.K.\ and S.C.O.G.\ acknowledge financial support from the ERC via Synergy Grant ``ECOGAL'' (project ID 855130) and from the German Excellence Strategy via the Heidelberg Cluster ``STRUCTURES'' (EXC 2181 - 390900948). In addition R.S.K.\ is grateful for funding from BMWE in project ``MAINN'' (funding ID 50OO2206), and from DFG and ANR for project ``STARCLUSTERS'' (funding ID KL 1358/22-1).
This paper makes use of the following ALMA data: ADS/JAO.ALMA \#2011.1.00172.S, ADS/JAO.ALMA \#2012.1.00108.S, ADS/JAO.ALMA \#2013.1.00191.S, ADS/JAO.ALMA \#2015.1.00274.S, ADS/JAO.ALMA \#2017.1.00895.S, and ADS/JAO.ALMA \#2018.1.00294.S. ALMA is a partnership of ESO (representing its member states), NSF (USA) and NINS (Japan), together with NRC (Canada), NSTC and ASIAA (Taiwan), and KASI (Republic of Korea), in cooperation with the Republic of Chile. The Joint ALMA Observatory is operated by ESO, AUI/NRAO and NAOJ. This research has made use of NASA’s Astrophysics Data System Bibliographic Services.

\bibliography{mybib}{}
\bibliographystyle{aasjournalv7}

\end{document}